\documentclass[prl,twocolumn]{revtex4-2}
\usepackage{epsfig}
\usepackage{graphicx}
\usepackage{amsfonts,amssymb}
\usepackage{amsmath}
\usepackage{color}
\usepackage{times}
\usepackage{flushend}
\usepackage{xcolor}
\usepackage{hyperref}

\hypersetup{
    colorlinks=true, 
    linkcolor=blue, 
    citecolor=blue, 
    filecolor=magenta, 
    urlcolor=blue, 
}

\begin{document}

\title{Distinct charge and spin recovery dynamics 
in a photo-excited Mott insulator}

\author{Sankha Subhra Bakshi and Pinaki Majumdar}

\affiliation{Harish-Chandra Research Institute
(A CI of Homi Bhabha National Institute), 
Chhatnag Road, Jhusi, Allahabad 211019
}
\pacs{75.47.Lx}
\date{\today}

\begin{abstract}
Pump-probe response of the spin-orbit coupled Mott insulator Sr$_2$IrO$_4$ 
reveals a rapid creation of low energy optical weight and suppression of 
three dimensional magnetic order on laser pumping. Post pump there is a quick 
reduction of the optical weight but a very slow recovery of the magnetic order 
- the difference is attributed to weak inter-layer exchange in Sr$_2$IrO$_4$ 
delaying the recovery of three dimensional magnetic order. We 
suggest that 
the effect has a very different and more fundamental origin. Combining  
spatio-temporal mean field dynamics and Langevin dynamics on the photoexcited 
Mott-Hubbard insulator we show that the timescale difference is not a dimensional 
effect but is intrinsic to charge dynamics versus order reconstruction in a 
correlated system. In two dimensions itself we obtain a short, almost pump fluence 
independent, timescale for charge dynamics while recovery time of magnetic order 
involves domain growth and increases rapidly with fluence. Apart from 
addressing
the iridate Mott problem our approach can be used to analyse phase competition 
and spatial ordering in superconductors and charge ordered systems out of 
equilibrium.
\end{abstract}

\maketitle


With the discovery of high temperature superconductivity in the doped
cuprates their parent Mott insulating state, for example La$_2$CuO$_4$
\cite{Grant},
has been extensively studied \cite{lee-nag-rmp}. 
The Mott state arises due to strong local repulsion in atomic orbitals,
which prevents simultaneous occupancy of both spin states, leading 
to an insulator at half-filling \cite{Imada,Perdew,Anisimov}. 
The localised electrons have an inter-site exchange interaction
that promotes antiferromagnetic order in non frustrated lattices.

When the Mott insulator is excited by a laser pulse with 
appropriately chosen frequency the added energy has an 
impact on both the charge dynamics and the  magnetic order
\cite{review1,review2,review3,ppmott1,ppmott2,ppmott3,ppmott4,
ppmott5,ppmott6,ppmott7,ppmott8,ppmott9,ppmott10,
ppmott11,ppmott12,ppmott13}.
The pump can excite 
electrons from the lower Hubbard band (LHB)
 to the upper Hubbard band (UHB),
which in real space means creation of double occupancy
\cite{Zala,Huang,Wrobel}.
The effect is twofold: (i)~the electrons excited to
the UHB, and the `holes' in the LHB, 
act as mobile carriers, leading to
metallic response in the optical conductivity,
and (ii)~the double occupancy suppresses the magnetic moments,
destroys their spatial correlation, and leads to suppression
of magnetic long range order.
In effect, despite being at half filling, one obtains a
transient metallic state coexisting with (small) local
moments, which evolves back 
towards its reference antiferromagnetic  
Mott state. 
It is in this context that Sr$_2$IrO$_4$ has thrown
up several puzzles about charge and spin dynamics
out of equilibrium.

A photo-excited state was realised in 
Sr$_2$IrO$_4$, a structural analog of La$_2$CuO$_4$.
At equilibrium Sr$_2$IrO$_4$ is a spin-orbit 
coupled layered antiferromagnetic Mott insulator 
(AFMI) with an interaction split $J_{eff} = 1/2$ band
\cite{material1,material2,material3,material4}.
The AF $T_c$ is $\sim 240$K.
Laser pumping leads to dramatic changes
in its  magnetic and electronic properties.
The two most significant observations in our reading
are: (i)~The rapid loss of 3D magnetic order and gain
in reflectivity in response to a pump - and then a 
 quick suppression of the reflectivity, within 1-2
picoseconds (ps) but 
very slow recovery of order (touching 1000 ps)
\cite{experiment1}.
This is the `two timescale' issue. 
(ii)~The persistence 
of low energy spectral weight in the optical conductivity
even after a seeming steady state is quickly 
reached, suggesting a population of excited high energy 
electrons - the holon-doublon
(HD) plasma  - at long times \cite{experiment2}. 
The time dependence of planar magnetic fluctuations
\cite{experiment1}, oddly, 
reflects both short and long timescales!
In the current interpretation
\cite{experiment1} the in plane
magnetic order and electron physics in the Mott insulator 
both recover quickly and the 
3D recovery is delayed due to weak interlayer 
coupling.

We suggest
that
the fascinating data revealed by the experiments has
another - very different - explanation.
This has remained out of reach because
theories of nonequilibrium phenomena need to consider 
the excited electronic population and also the spatial 
dynamics, e.g, domain growth effects, on large spatial
scales.  
Current tools, e.g, exact diagonalisation 
\cite{ED1,ED2}, dynamical mean field theory (DMFT) 
\cite{DMFT1,DMFT2,DMFT3,DMFT4,DMFT5,DMFT6}.
and density matrix renormalisation group
\cite{DMRG1,DMRG2}, are either size limited or
unable to access the timescale needed. 

To capture the electron correlation non perturbatively and
access spatio-temporal dynamics we write an
equation of motion for the one body density operator 
$ \rho_{ij}^{\sigma \sigma'} = 
c^{\dagger}_{i\sigma} c_{j \sigma'}$ in the
half-filled square lattice
Hubbard model and close the hierarchy by factorising the 
resulting four operator terms in the magnetic channel.  
This is `mean field dynamics' (MFD) although
it is in terms of the matrix $\rho_{ij}^{\sigma \sigma'}$
rather than a local object like magnetisation.
This allows access to size $\sim 20
\times 20 $ and time upto  $10^{3}t^{-1}_{hop}$,
where $t_{hop}$ is the in-plane hopping. 
We also construct, benchmark, and use a Langevin 
dynamics (LD) scheme that allows access to $\sim 60 \times
60$ lattices and time upto $10^{4}t^{-1}_{hop}$. 
We set $t_{hop} = 260$meV ($t_{hop}^{-1} \equiv 16$ femtoseconds) 
and $U = 3 t_{hop}$ as noted
\cite{material1} for Sr$_2$IrO$_4$. 
Our results, based on a 
combination of MFD and LD,
are the following.

(i)~The pump induced suppression of order 
and enhancement of optical weight occurs over $\sim 0.3$ps. 
Post pump, the optical weight reduces to reach its long time 
asymptote over  $\tau_{opt} \sim 2-3$ ps, 
while magnetic order recovery time $\tau_{ord}$ 
ranges from a few ps to $100$ps depending 
on fluence.

(ii)~The `time resolved' magnetic fluctuation spectrum 
$S({\bf q}, \omega, t)$ has a recovery time $\tau_{fluc}({\bf q})$
that varies widely,
$\tau_{fluc}({\bf q}) \sim \tau_{opt}$ when ${\bf q}$ is far from 
the ordering vector ${\bf Q} = (\pi, \pi)$,
 and $\tau_{fluc}({\bf q}) \sim \tau_{ord}$ when 
${\bf q} \rightarrow (\pi, \pi)$.

(iii)~While `charge recovery' to a steady state value,
seen in $\tau_{opt}$, is quick, a significant UHB  
electron population survives to long time, sustaining 
low frequency optical weight.
$\tau_{opt}$ and the weight correlates with measurements
\cite{experiment2}.

(iv)~Studying a layered $O(3)$ symmetric model with 
$J_{\perp}/J_{\parallel} \sim 10^{-3}$ we found that 
the 3D `recovery time' differs from the 2D value by 
only a factor $\sim 2$. Key dynamical features of the
layered material are two dimensional.

{\it Model:}
Although Sr$_2$IrO$_4$  is a three dimensional multiband 
system with spin-orbit coupling the essential physics is 
in the IrO planes, 
controlled by a single spin-orbit coupled  
orbital per site subject to a Hubbard 
interaction. The single band model takes the conventional
form: $ H=\sum_{\langle ij \rangle,\sigma} 
t_{ij}c^{\dagger}_{i\sigma}c_{j\sigma}
+ U \sum_{i} n_{i\uparrow}n_{i\downarrow} $.
We set the lattice spacing to 1.

We introduce the pump via a pulse of 
amplitude $E_0$, 
frequency 
$\Omega_p \sim 120$~THz and  
pulse width $\tau_p=100$fs 
by Peierls coupling. 
$E_0^2$ would be proportional to the fluence.
The evolution of 
$\langle \rho_{ij}^{\sigma \sigma'} \rangle$
involves solution of the $4N^2$ MFD 
dynamical equations  \cite{MFD},
where $N$ is the number of sites.
The form of this equation, its derivation, 
and its implementation is
discussed in Supplement A \cite{suppl}, and
primary indicators shown in Supplement B. 
The local magnetisation can be calculated as 
${\vec m}_i =
\sum_{\sigma,\sigma'} \vec{\tau}_{\sigma\sigma'}
\langle \rho_{ii}^{\sigma \sigma'} \rangle$, where
$\tau$ are the Pauli matrices.

While MFD reveals key features of 
`suppression-recovery' dynamics in response to the pump, 
it has limitations in terms of accessible size and time
that prevent study of the timescale separation observed 
in experiments.
Electronic properties, like the upper Hubbard band occupancy,
stabilize to a pump-dependent constant value on a short 
timescale 
$t \sim \tau_{opt} \sim 2$ ps, as confirmed within MFD to 
$t \sim 10 \tau_{opt}$. 
Beyond a few ps, the population of electrons excited to 
the upper 
Hubbard band stabilizes to a finite time-independent value, 
akin to having a steady finite 'electronic temperature' 
$T_{el}$ at long times. 
This constancy is then used in Langevin calculations 
on large lattices 
for times up to $(50-60) \tau_{opt}$. The detailed 
characterization of 
$T_{el}(t,E_0)$ is provided in Supplement B, where 
$T_{el}$ serves as a 
crucial input for modeling magnetic dynamics below.

To address the magnetic dynamics 
with high spatial resolution and long times
we write a Langevin equation 
directly for the  ${\vec m}_i$. 
In such an equation 
the  ${\vec m}_i$ are subject 
to a torque arising from the electrons,
a damping, and
a `thermal' noise at the bath temperature $T_b$.
The pump excited electrons 
follow a Fermi distribution 
with temperature $T_{el}$ inferred from MFD.
The equation takes the form:
\begin{eqnarray}
{ {d {\vec m}_i} \over {dt}}  &=&
-\vec{m}_i \times \frac{\partial \langle F_{SF} \rangle}
{\partial \vec{m}_i} - \gamma \frac{\partial
 \langle F_{SF} \rangle}
{\partial \vec{m}_i}+ \vec{\eta}_i
\cr
\cr
\langle \eta_i^{\alpha}(t) \rangle &=&0, ~~~
\langle \eta_i^{\alpha}(t) \eta_j^{\beta}(t')\rangle =
 2 \gamma k_BT_{b}
\delta_{ij} \delta_{\alpha \beta} \delta(t - t')
\nonumber
\end{eqnarray}
$F_{SF}$ is the free energy of electrons at a 
temperature $T_{el}$ in
the spin background $\{{\vec m}_i\}$, $\gamma$ is a
dissipation constant extracted from MFD, and $\eta_i$ is
thermal noise with $\eta$ and $\gamma$ satisfying the
fluctuation-dissipation theorem at temperature $T_b$.
The free energy $F_{SF}$ arises from the 
Hubbard model,  $H_{SF}$, written in the
spin-fermion language:
\begin{eqnarray}
H_{SF}\{{\vec m}\} 
& =& \sum_{ij,\sigma} t_{ij}c^{\dagger}_{i,\sigma} c_{j,\sigma}
- 2U\sum_{i}{\vec{m}}_i.{\vec{s}}_i + U\sum_{i}\hat{\vec{m}}_i^2
\cr
\cr
&=& \sum_{n} \epsilon_n f^{\dagger}_n f_n + 
U\sum_{i}\hat{\vec{m}}_i^2
\cr
\cr
F_{SF}\{{\vec m}\} &=&
- T_{el} \sum_n ln(1 + e^{-\beta_{el} (\epsilon_n
-\mu)}) +  U\sum_{i}{\vec{m}}_i^2
\nonumber
\end{eqnarray}
${\vec s}_i = c^{\dagger}_{i \sigma}  {\vec \tau}_{\sigma \sigma'}
c_{i \sigma'}$ is the electron spin operator.
$\epsilon_n$ in the last line are the 
single particle eigenvalues for the electron system in a spin 
background $\{{\vec m}_i\}$, and $\mu = U/2$.
\begin{figure}[b]
\centerline{
\includegraphics[width=4.2cm,height=3.4cm]{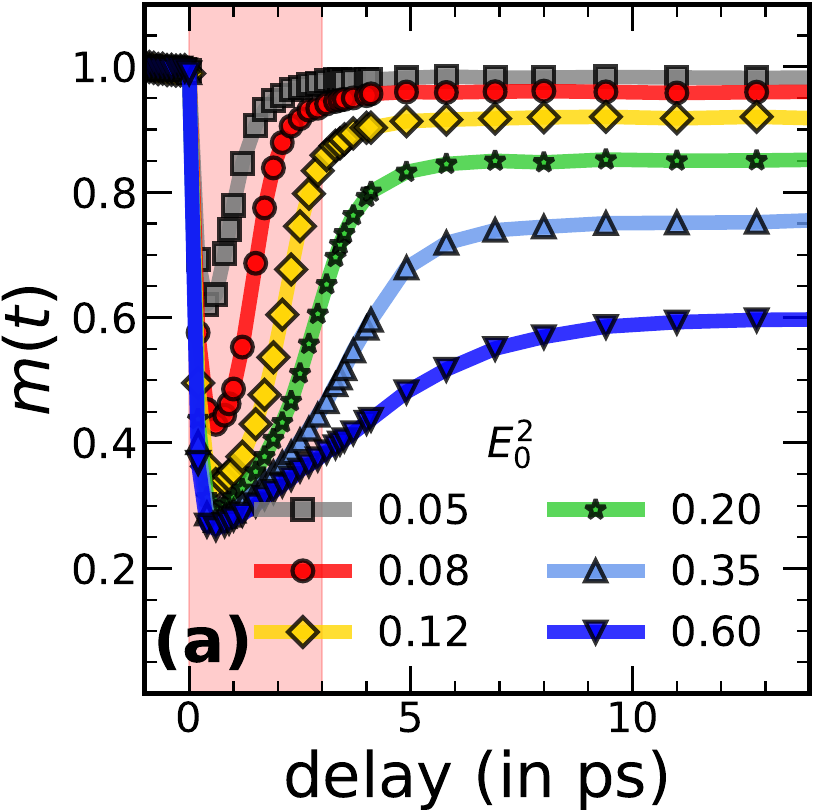}
\includegraphics[width=4.2cm,height=3.4cm]{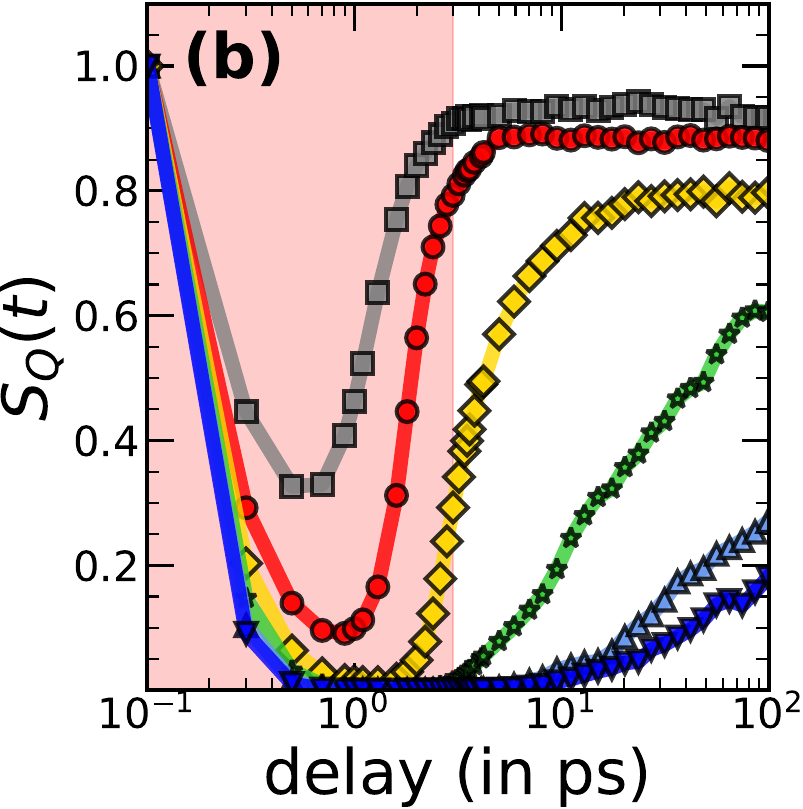}
~
}
\centerline{
\includegraphics[width=8.2cm,height=3.0cm]{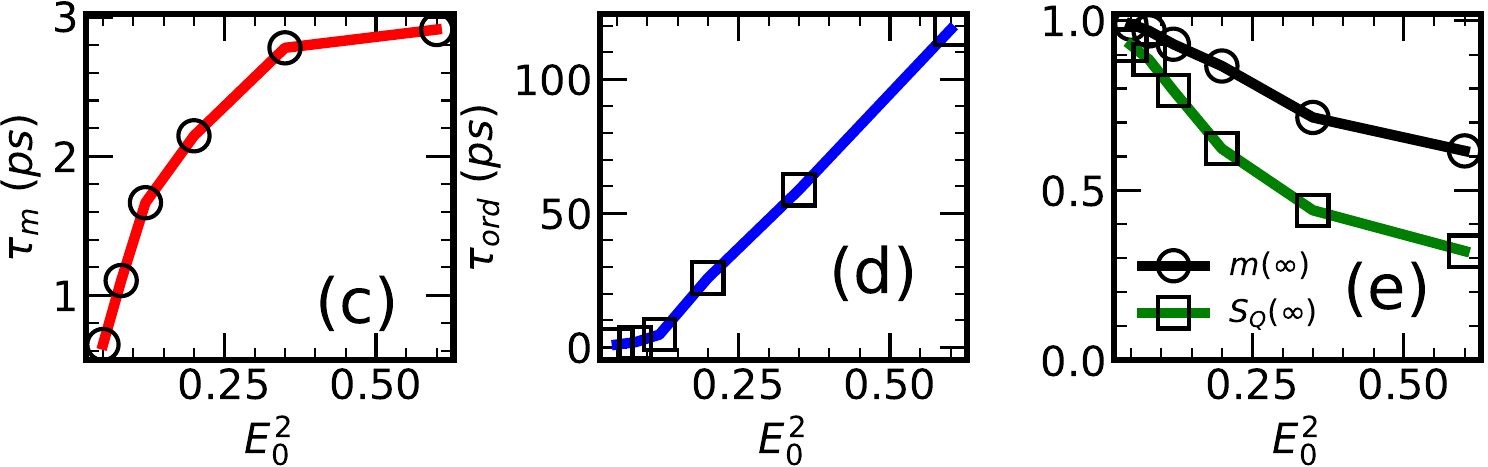}
~~
}
\caption{Mean magnetic moment and 2D magnetic order. 
(a)~The instantaneous system averaged magnetic moment,
$m(t)$, normalised to $1$ at $t=0$. Following an abrupt 
increase in electron temperature from $T_{b}$ to $T_{el}^i$ 
there is a quick decrease in $m(t)$. As $T_{el}$ decreases 
towards $T_{el}^f$,  $m(t)$ recovers towards a smaller
final value. Relation between $T_{el}^f$ and $E_0^2$ is 
established in Supplement. 
(b)~$S_{\bf Q}(t)$ is initially sharply 
reduced on pulse impact and grows slowly even 
after $m(t)$ reaches a steady value.  
(c)~Revival time 
for $m(t)$: the maximum value is $\sim 3$ps at the largest 
$E_0^2$.  (d)~Revival time for 2D magnetic order is upto 
$120$ps.  (e)~The dependence  of $m(\infty)$ and
$S_{Q}(\infty)$ on $E_0^2$.
}
\end{figure}
Calculating 
${\partial F_{SF}}/{\partial \vec{m}_i}$, which
is $\langle {\vec s}_i \rangle$, requires knowledge
of the eigenvalues and eigenfunctions of
the whole system.  At $U/t_{hop}=3$ 
we can calculate  $\langle {\vec s}_i \rangle$ 
accurately by constructing a cluster around the site ${\bf R}_i$
and diagonalising the cluster Hamiltonian instead of
having to diagonalise the full $H$. We use a 
13 site cluster centred on ${\bf R}_i$, 
including 4 nearest neighbour sites and 8 next nearest
neighbour sites.
We have benchmarked this against the full 
diagonalisation based calculation.

The electron temperature profile obtained from MFD
can be approximated as:
$
T_{el}(t) = T_{el}^ie^{-t/\tau_{el}} + 
T_{el}^f (1 - e^{-t/\tau_{el}}) $.
We find $\tau_{el} \sim 2 $ ps and
$T_{el}^f \propto T_{el}^i$ and a ratio
$T_{el}^i/T_{el}^f = 2.5$ allows us to model all $E_0$
using only one $\gamma$ value 
(see Supplement C).
This leave $T_{el}^f$ as the only
fluence dependent parameter in the 
Langevin equation.
The fit $T_{el}^f(E_0)$ is 
shown in the Supplement, where we also
estimate $\gamma \approx 0.1$. 
We set bath temperature $T_b=40$K, the temperature for 
experiments were $\sim 80-100$K. 
We use the 
Euler-Maruyama algorithm to solve  the LD equation  
with step size $\delta t \sim 0.01 t_{hop}^{-1}$. 
We average the Langevin data over 
5-10 runs when extracting timescales
for the magnetic response. 

{\it Timescales:}
In Fig.1 we show the time dependence of the system averaged
magnetic moment $m(t)=\frac{1}{N}\sum_i |\vec{m}_i|$ and the 2D structure factor 
$S_{\bf Q}(t)=\frac{1}{N^2} \sum_{ij} 
e^{i {\bf  Q}.({\bf R}_i - {\bf R}_j)}
{\vec m}_i (t)\cdot{\vec m}_j (t)
$ at ${\bf Q} = (\pi, \pi)$.
We set the pre-pulse values of
$S_{\bf Q}$ and $m$ to 1. As the pulse hits 
the system both $m(t)$ and $S_{\bf Q}(t)$ 
are suppressed on a timescale $\sim 0.3$ps.
Post pulse, the mean moment rises quickly (panel (a))
while $S_{\bf Q}$ (panel (b)) has a strongly fluence
dependent recovery time. The time axis in panel (b) is
logarithmic, highlighting the wide range of recovery times.
We fit the suppression-recovery dynamics in $m(t)$ and
$S_{\bf Q}(t)$ to simple exponentials of the form: 
$
I(t) = I(0) 
e^{-t/\tau_d} + I(\infty) (1-e^{-t/\tau_r})
$
For both $m$ and $S_{\bf Q}$ we find  
$\tau_{d}\sim 0.2$ ps. Panel (c) shows the
recovery time $\tau_m$ of $m(t)$, while panel (d)
shows the recovery time $\tau_{ord}$ for $S_{\bf Q}$.
$\tau_m$ ranges from $1-3$ps while $\tau_{ord} $ 
ranges from $1-120$ps. Size dependence of $\tau_{ord}$
is shown in Supplement D. 

The associated long term amplitudes  $m(\infty)$ 
and $S_{\bf Q}(\infty)$ are
shown in panel (e). For an experimental system in a 
thermal environment, $m(t)$ and 
$S_{\bf Q}(t)$ should return to their pre-pump value 
after the post pump system attains equilibrium. 
Within our MFD this does not happen since a fraction
of the pump created  double occupancy persists
at long time. Their deexcitation requires multimagnon
emission processes which have a long timescale and
are beyond MFD.
Even in the experiments some indicators
do not return to pre-pump values over experimental
observation time. 

\begin{figure}[b]
\centerline{
\includegraphics[width=8.5cm,height=4.4cm]{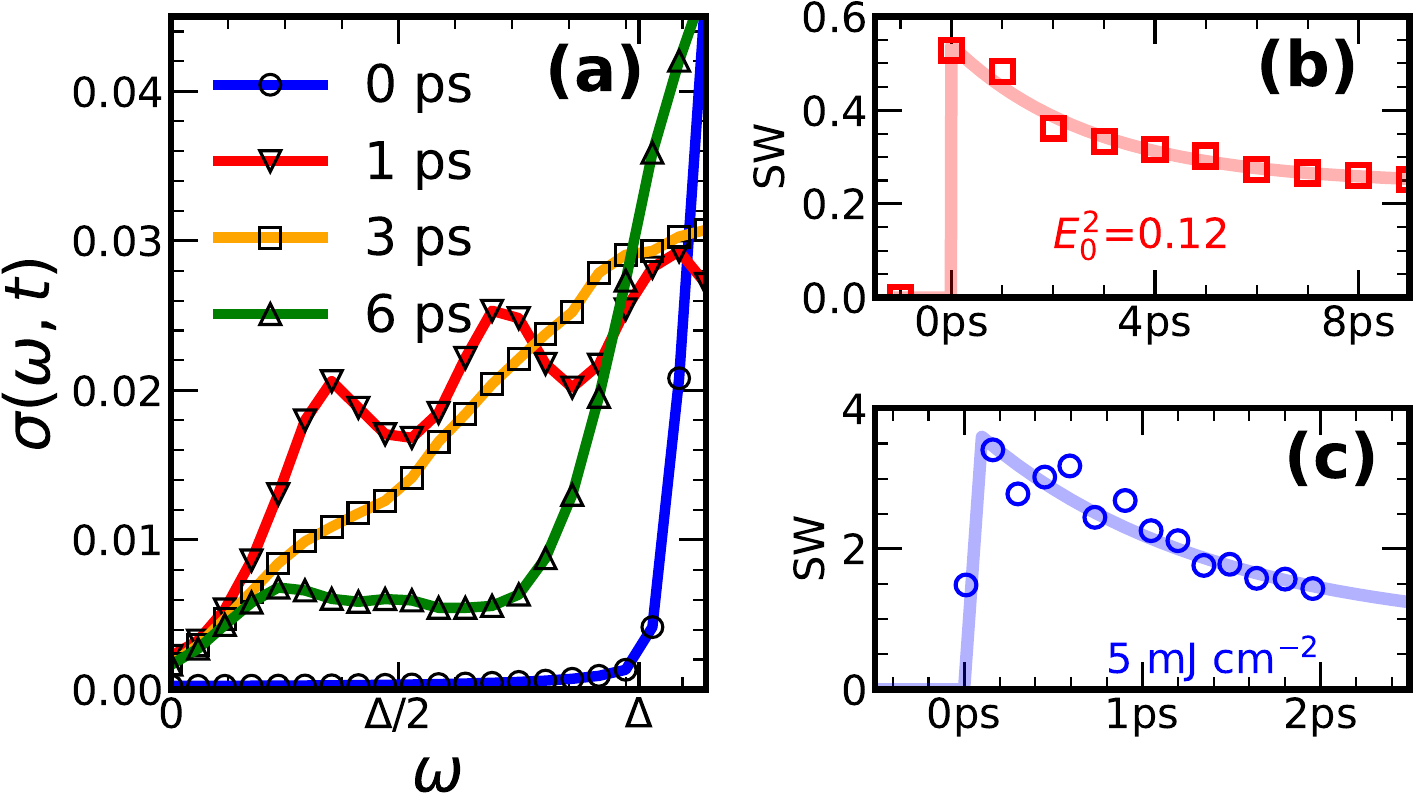}
~~}
\caption{Time dependence of low-frequency optical
conductivity and spectral weight.
(a)~$\sigma(\omega)$ over $\omega =  [0 - \Delta]$
is plotted for $E_0^2 = 0.12$ at several post pulse
times. At equilibrium $\sigma_R(\omega) = 0$ for
$\omega < \Delta$.
The excited, and slowly decaying, upper band
population leads to the time dependence shown.
(b)~Time dependence of the spectral weight (SW) 
integrated over $\omega =  [0.05 \Delta - 0.1 \Delta]$.
The SW initially
rises from zero, then decays, and stabilizes at a finite value.
(c)~Experimental data on time dependence of post pump
 low energy SW. Integrated over $1-2.5$THz for 
pump fluence of 5 mJ cm$^{-2}$.
Faint lines in (b, c) are fit with an exponential decay. 
}
\end{figure}
\begin{figure}[t]
\includegraphics[width=8.8cm,height=8.0cm]{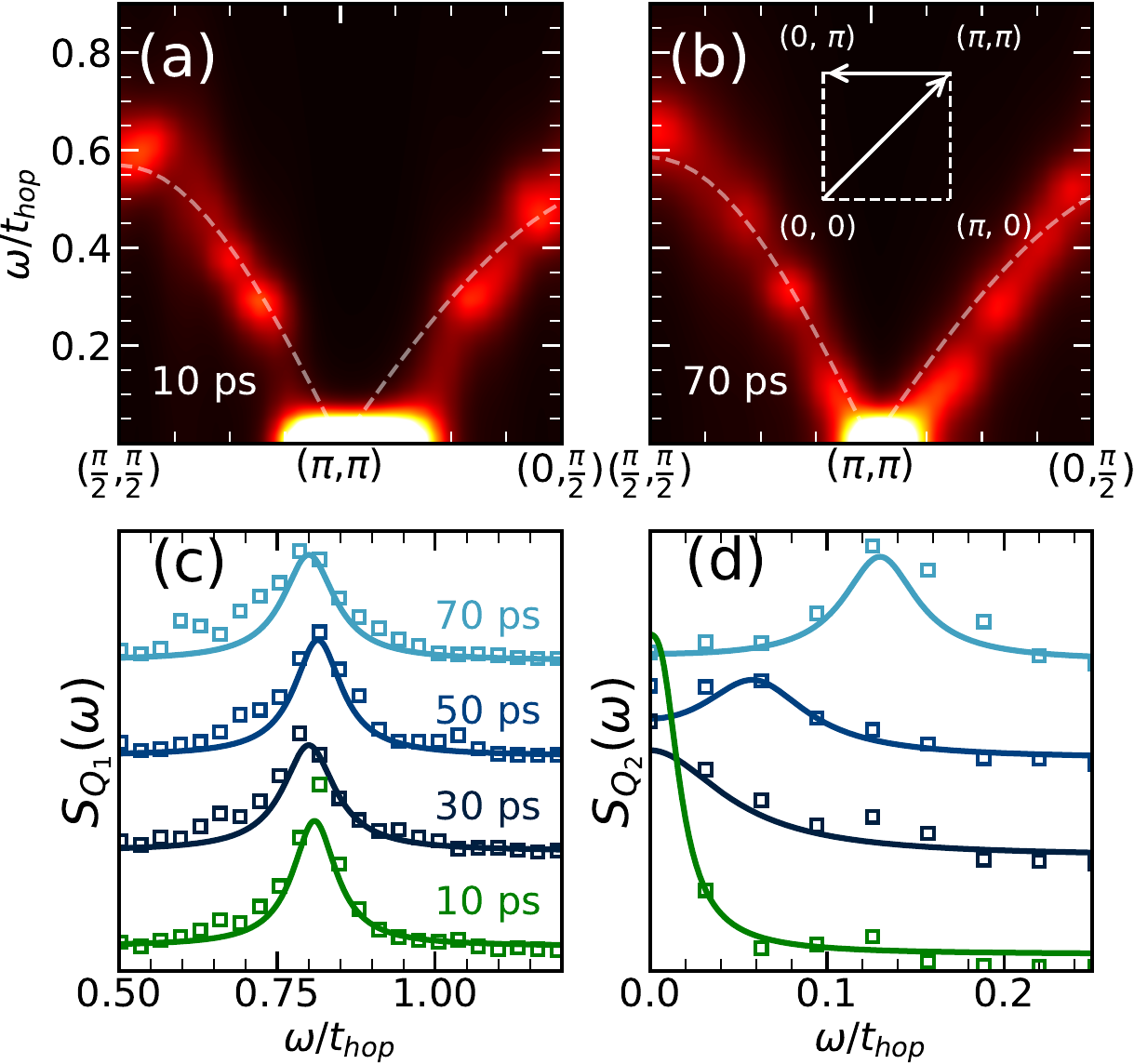}
\caption{Dynamical structure factor $S({\bf q},\omega,t)$
for $E_0^2 = 0.4$
with a time window of $\pm 10$ps centered at $t$.
(a)-(b)~Momentum scans of $S({\bf q},\omega,t)$ for
$t=10,~70$ps, averaged over five LD runs.
The equilibrium spin wave dispersion $\omega_0({\bf q})$
is superposed.
We see a dramatic deviation from $\omega_0({\bf q})$ at 10ps 
for ${\bf q} \sim {\bf Q}$. The spectrum mostly 
recovers towards
$\omega_0({\bf q})$ by 70ps, except near ${\bf Q}$.
(c)~Time dependence of the lineshape 
at ${\bf Q}_1 = (0, \pi)$ 
and (d)~at ${\bf Q}_2 = (0.9\pi, 0.9\pi)$.
The spectrum at~${\bf Q}_1$ is stabilised by 10ps,
while the spectrum at~${\bf Q}_2$ is evolving even at 70ps.
}
\end{figure}

{\it Optical response:}
The most dramatic effect, and readily measurable 
consequence, of the excited electron population
is in the  optical conductivity $\sigma(\omega)$. 
At equilibrium the Mott
insulator at low temperature should have no weight in
the real part of $\sigma(\omega)$, for $\omega < \Delta$,
the equilibrium gap.
Since the electronic timescale 
is $\sim $ ten times shorter than magnetic timescale
(as inferred from the magnetic bandwidth in Fig.3, later), 
we calculate \(\sigma(\omega,t)\) using the instantaneous 
electronic eigenstates and eigenvalues at time \(t\) 
(see Supplement E).

Fig.2 shows features of $\sigma(\omega,t)$ at
$E_0^2 = 0.12$, the fluence at which $S_{\bf Q}$ first
drops to zero, Fig.1.(b).  In Fig.2(a) $\sigma_R(\omega)$
at $t=0$ shows the absence of
any weight for $\omega < \Delta$. Between $t=0$ and $1$ps
the  weight in the interval $\omega = [0,\Delta]$ 
rises quickly, reaching a maximum around 1ps
and declining thereafter.
\(\sigma(\omega)\) deviates from 
a Drude response due to the suppressed low energy
density of states and the strong orientational
disorder in the magnetic background.

In panel (b) we show the integrated weight over a small 
window $\omega = [0.05 \Delta - 0.1 \Delta]$,  as had been
done in the pump probe experiment \cite{experiment2}.
The two noteworthy features are (i)~the quick decay to
a long time value with a time constant 
$\tau_{opt} \sim 2$ps, (ii)~a `long 
time' value that is $\sim 50 \%$ of the peak. 
Panel (c) shows the time dependence of the weight 
extracted from the experimental data in Fig.2(g) 
in \cite{experiment2}. 

It is not coincidental that the optical timescale
$\tau_{opt}$ and the moment recovery time $\tau_m$
are comparable. Both arise from the deexcitation of
electrons pushed to the upper band by the pump pulse.
The deexcitation reduces double occupancy, trying
to restore the moment magnitude, and the reducing 
number of electrons in the UHB, and `holes' in the
LHB, reduces the low energy optical weight. 
Broadly speaking, these process are related to
the quick - local - 
charge relaxation in the system, operative
on a few ps timescale. 
This contrasts with the long timescale for
restoring global order.
We next look at an indicator - already probed
experimentally - of the momentum and 
`time resolved' magnetic fluctuation spectrum.

{\it Spin dynamics:}
Experiments have measured the 2D dynamical 
structure factor $S({\bf q},
\omega)$ of the spins via time resolved 
resonant inelastic X-ray scattering (tr-RIXS)
\cite{experiment1}.
In contrast to the equilibrium case this collects 
`${\bf q}-\omega$' data of magnetic fluctuations over a
time interval $\pm \Delta t$ around a reference time $t$.
The background state is time evolving so the resulting
$S({\bf q}, \omega)$  depends on the reference time $t$.
We compute the corresponding object based on LD data:
$
S({\bf q}, \omega, t) 
= {1 \over N^2} {\Big \vert} \sum_i \int_{t - \Delta t}^{t + \Delta t}  dt'
e^{i {\bf q}.{\bf R}_i} e^{i\omega t'} {\vec m}_i(t') {\Big \vert}^2 
$
We set $\Delta t = 10$ps and used $t=10,30,50,70$ps.

\begin{figure}[t]
\includegraphics[width=8.8cm,height=5.3cm]{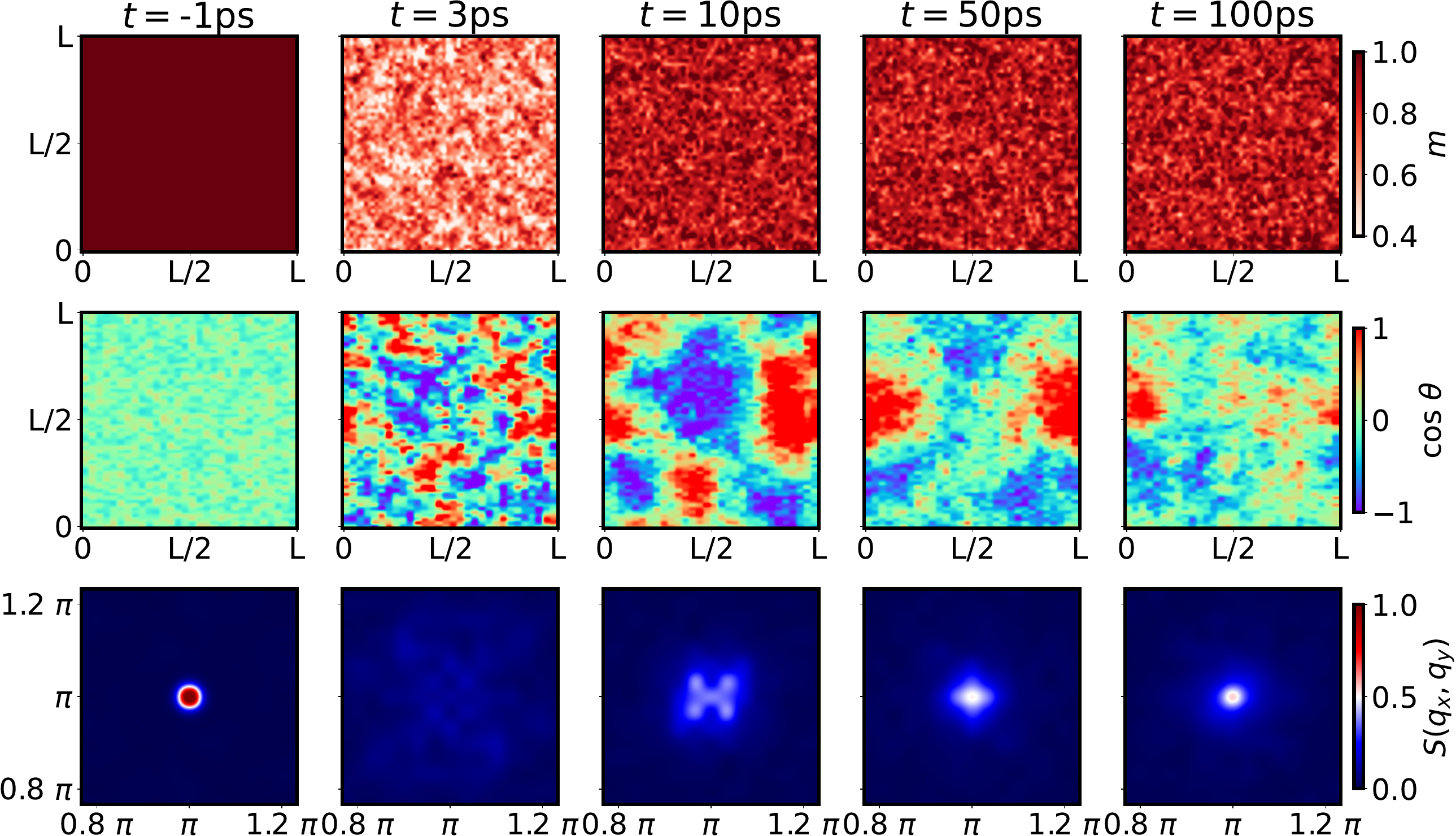}
\caption{Time resolved spatial maps at $E_0^2=0.4$
on a $60\times 60$ lattice.
Top row: spatial variation of local moment size $|vec{m}_i|$.
The uniform pre-pulse magnitide is strongly suppressed and
randomised at $t=3$ps but attains a steady state character by
$10$ps.
Middle row: locally AF ordered domains constructed out of the
orientation of the local moments ${\vec m}_i$. The
single domain pre-pulse state is fragmented into tiny domains
at $3$ps and gradually evolves towards a single domain by $t=100$ps.
Bottom row:
Structure factor $S_{q_x,q_y}$, evolving from the pre-pulse 
single peak at ${\bf q} = {\bf Q}$ to a featureless form
at $t=3$ps, and then the gradual re-emergence of weight near
 ${\bf q} = {\bf Q}$. A clear peak at ${\bf Q}$ arises
only at $t=50$ps before which competing domains cancel off
any system wide order.
}
\end{figure}

Fig.3(a)-(b) show $S({\bf q}, \omega)$ as colour maps for
two values of reference time, $10$ps and $70$ps.
On the  $x$ axis is ${\bf q}$, on the $y$ axis is $\omega$,
and the intensity is coded in colour. 
The results are at 
$E_0^2 = 0.4$, a `strong fluence' case where magnetic order
recovers very slowly. 
Superposed on (a) and (b) is the equilibrium spin wave
dispersion $\omega_0({\bf q})$  of the $(\pi,\pi)$ AF state. 
At both $t$ the spectrum far from the ordering
wavevector $(\pi,\pi)$ looks similar to $\omega_0({\bf q})$
except for a correction due to moment size. However
near $(\pi,\pi)$ the spectral weight distribution 
is very different from $\omega_0({\bf q})$: at $t=10$ps
the difference is drastic while at $t=70$ps $S({\bf q}, \omega)$
still has not recovered its character. Note that these times
are much greater than $\tau_m$ or $\tau_{opt}$ that we
have discussed.

To highlight the ${\bf q}$ dependence of the recovery process
panels (c) and (d) 
show the lineshapes for two momenta: one situated 
far from $(\pi, \pi)$ at $(0, \pi)$, the other 
near $(\pi, \pi)$ at $(0.9\pi, 0.9\pi)$,
In each of these we have highlighted $\omega = \omega_0({\bf q})$
by an arrow. 
It is obvious that for  $(0, \pi)$  the peak location in 
 $S({\bf q}, \omega)$ has stabilised within $10$ps. 
At $(0.9\pi, 0.9\pi)$ however the peak location
is far from stabilised even
at $70$ps. This is not surprising
given that the ordering 
timescale $\tau_{ord}$ at this fluence is
$\sim 60-70$ps, Fig.1(d), and it would take $\sim 2-3$~ $\tau_{ord}$
for the spectrum near ${\bf Q}$ to stabilise.

More generally, the results in Fig.3 suggest that the `recovery'
time for the magnetic fluctuation spectrum is strongly ${\bf q}$
dependent, and this timescale $\tau_{fluc}({\bf q})$ ranges
from $\tau_m$ when ${\bf q}$ is far from ${\bf Q}$ to $\tau_{ord}$ 
when ${\bf q} \rightarrow {\bf Q}$.
The RIXS spectrum probes moment size recovery and short
range correlations when far from ${\bf Q}$, and progressively
longer range correlation of the moments when ${\bf q} \rightarrow 
{\bf Q}$.  In the next figure we wish
to show the `domain growth' physics that underlies this
phenomena.

{\it Spatial behaviour:}
Fig.4 shows the detailed spatial behaviour of the
moment size $\vec{m}_i$ in the top row, differently
oriented AF domains (coded by colour) in the middle row,
and the 
instantaneous structure factor $S_{\bf q}$, in the 
bottom row, for time ranging from $-1$ps to $100$ps.
This is a strong pulse situation, the same as studied
for $S({\bf q}, \omega)$.

As the pulse hits the system the mean magnetic moment reduces 
to 40$\%$ of its original value and then quickly recovers 
to a stable value by 10ps. There is no spatial structure to the 
fluctuations in $\vec{m}_i$. The middle row shows 
the pre-pulse perfect order at $t=-1$ps - a single AF domain.
At $3$ps, where the moment value is still suppressed, 
there is a patchwork of small domains with linear dimension of a
few lattice spacings (the map is $60 \times 60$). By $10$ ps 
the moments have stabilised and there are only 
a few large competing domains.
$50$ps and $100$ps show the increasing dominance
of one (green-blue) domain. 
The lowest row shows the ${\bf q}$ dependence of $S_{\bf q}$
near the ordering wavevector. The perfect order in the
pre-pulse state is destroyed at 3ps due to suppression of the moments.
Then there is a slow growth of intensity near  ${\bf q} = {\bf Q}$
and a clear peak becomes visible only at $50$ps. 

In the domain pictures we have gauged out the $(\pi, \pi)$ 
oscillation of Neel order and plotted equivalent
`ferromagnetic' domains with net moment pointing in different 
directions.
If a domain has linear dimension $L_d$, probing it should 
lead to a fluctuation spectrum mimicking the bulk order as
long
as $q_x, q_y \ll \pi/L_d$. The typical $L_d$ for us is
time dependent, so the moderate size domains at
$10$ps well capture the fluctuation at $(\pi, 0)$, while
we would need $t \gtrsim 100$ps to capture
fluctuations near $(\pi, \pi)$. 

We discuss two issues now that have bearing on the
theory-experiment comparison and the reliability of the
calculation itself.
(i)~{\it 2D versus 3D:} The experimental paper
\cite{experiment1} suggested
that the growth of 3D order was slow because the interplanar
magnetic exchange $J_{\perp}$ was $\ll J_{\vert \vert}$, the
in plane value. It was implicit that recovery of 2D order 
was quick and the delayed 3D recovery was a $J_{\perp}/J_{\vert \vert}$
effect. We have shown that 2D recovery itself is inevitably
delayed. Does that mean an even more delayed 3D recovery?
We did dynamics on a `layered' Heisenberg model (Supplement F) 
with $J_{\perp}/J_{\vert \vert}$ down to $10^{-3}$. Naively one
may have expected $\tau_{3D}/\tau_{2D} \sim 10^{3}$.
We find the ratio to be 2!  This is 
because 2D is the lower critical dimension for $O(3)$
models and any small $J_{\perp}$ is a singular perturbation
\cite{heis1,heis2}.
(ii)~{\it Thermalisation:}
Within the MFD framework, from which we extract our $T_{el}(t)$,
there are pump induced holon-doublon excitations which
persist to
long time, and their effective temperature differs from 
the apparent temperature sensed by the magnetic moments. 
For a system that equilibriates, these two temperatures
should finally be the same. The electronic excitation scale $\Delta
\gg J_{\vert \vert}$ the magnetic excitation scale, so
multimagnon emission processes are needed to deexcite
electrons. The timescale arising from such processes has
been estimated to be 
$\sim t_{hop}^{-1} e^{ \alpha (\Delta/J_{\vert \vert} log(U/t_{hop})) }$,
\cite{therm}
where $\alpha \sim O(1)$. Plugging in $\Delta \sim 500$ meV and $J_{\vert \vert}
\sim 25$ meV, 
we would get a decay time $10^4$ ps, which is $\gg$ than the experimental observation time
or our run time.

{\it Conclusion:}
Pump-probe experiments have made it possible to 
temporarily `metallise' an antiferromagnetic Mott insulator
and suppress it's magnetic order by creating double occupancy
and destroying the magnetic moment. The key question was how
this strongly perturbed state evolves back towards the 
reference AFMI at long times. 
With the experimental data on Sr$_2$IrO$_4$ setting a 
reference we 
set up a hierarchical scheme that addresses this
nonequilibrium correlated problem on large spatial scales
in real time.  We conclude that 
the `charge physics', of optics etc, is dominantly local, 
quick,  and mostly insensitive to fluence. The magnetic 
order recovery, however, is strongly non local,
involves growth of domains, and brings in a fluence and
system size dependent timescale.
Momentum resolved magnetic excitations probe different
spatial scales, and hence different recovery times.
While our specific results are on the AFMI in Sr$_2$IrO$_4$,
the mean field dynamics framework, and it's reduced
Langevin counterpart, can be readily adapted to address
large spatial scale nonequilibrium phenomena in 
ordered systems like superconductors or charge density waves.
\\
\\
We acknowledge use of the HPC clusters at HRI. PM thanks Rajdeep
Sensarma for a discussion.


\newpage

\onecolumngrid
\begin{center}
    \textbf{\LARGE Supplementary to ``Distinct charge and spin recovery dynamics\\
    in a photo-excited Mott insulator''} \\[2em]
    \textbf{Sankha Subhra Bakshi and Pinaki Majumdar} \\[1em]
    \small Harish-Chandra Research Institute \\ 
    \small (A CI of Homi Bhabha National Institute), \\
    \small Chhatnag Road, Jhusi, Allahabad 211019
\end{center}

\section{Supplement A: Mean field dynamics (MFD)}

In this section we derive an equation of motion for the `density 
operator' ${\hat \rho}_{ij}^{\sigma \sigma'} = c^{\dagger}_{i\sigma}c_{j \sigma'}$,
 from which the local
magnetisation and local density can be computed.
We start with a single band repulsive Hubbard model
\begin{equation}
    H =  H_t + H_U  = \sum_{ij,\sigma} t_{ij}
    c^{\dagger}_{i,\sigma} c_{j,\sigma} 
    + U\sum_{i}\hat{n}_{i\uparrow}\hat{n}_{i\downarrow}
\end{equation}
Where $t_{ij}$ is the nearest neighbor hopping with value 
$t_{\text{hop}}=-1$.
We can rewrite the interaction term in the following way:
\begin{align}
U \sum_i c^\dagger_{i,\uparrow} c_{i,\uparrow} c^\dagger_{i,\downarrow}
 c_{i,\downarrow} = \frac{U}{4} 
\sum_i n_i^2  -U \sum_i (\Vec{s}_i.\hat{R})^2
\end{align}
Where the local density operator is 
$n_i =c^\dagger_{i\uparrow} c_{i\uparrow}
 + c^\dagger_{i\downarrow} c_{i\downarrow}$, 
and the local spin operator is $\Vec{s}_i = 
\sum_{\sigma, \sigma'} c^\dagger_{i\sigma} 
\Vec{\tau}_{\sigma\sigma'} c_{i\sigma'}$,  
$\Vec{\tau}$ being the Pauli matrices. 
$\hat{R}$ is any arbitrary SO(3) unit vector.
The interaction term can be rewritten as 
\begin{equation}
    U \sum_i c^\dagger_{i\uparrow} c_{i\uparrow} c^\dagger_{i\downarrow}
c_{i\downarrow} = \sum_{i} \sum_{\alpha \beta \alpha' \beta'}
    h_{\alpha \beta \alpha' \beta'} c^\dagger_{i\alpha} c_{i\beta} 
c^\dagger_{i\alpha'} c_{i\beta'} 
\end{equation}
where 
\begin{equation}
    h_{\alpha \beta \alpha' \beta'} 
    = \frac{1}{4} \delta_{\alpha\beta}\delta_{\alpha'\beta'}
    -(\vec{\tau}_{\alpha\beta}. \hat{R})
    (\vec{\tau}_{\alpha'\beta'}.\hat{R})
\end{equation}
Now we write the Heisenberg equation for 
$\hat{\rho}_{ij}^{\sigma\sigma'}(t)$
\begin{eqnarray}
{ {d\hat{\rho}_{ij}^{\sigma\sigma'}} \over {dt}}  &=& 
-i [\hat{\rho}_{ij}^{\sigma\sigma'}, H] 
\end{eqnarray}
The bilinear term in $H$ produces bilinear correlations
\begin{eqnarray}
    -i [\hat{\rho}_{ij}^{\sigma\sigma'}, H_t] &=&
    i \sum_k (t_{ki}\hat{\rho}^{\sigma\sigma'}_{kj}
    -\hat{\rho}^{\sigma\sigma'}_{ik}t_{jk})
\end{eqnarray}
The interaction term leads to:
\begin{eqnarray}
    -i [\hat{\rho}_{ij}^{\alpha\beta}, H_U] &=&
    2iU \sum_i \sum_{\alpha,\beta,\gamma}
    (h_{\sigma'\gamma\alpha\beta} 
    \hat{\rho}^{\sigma\gamma}_{ij}
    \hat{\rho}^{\alpha\beta}_{jj}
    -h_{\gamma\sigma\alpha\beta} 
    \hat{\rho}^{\gamma\sigma}_{ij}
    \hat{\rho}^{\alpha\beta}_{ii}
    )
\end{eqnarray}
We take average of both sides and write
$\langle\hat{\rho}^{\alpha\beta}_{ij}\rangle=\rho^{\alpha\beta}_{ij}$.
The interaction term produces a term of the form 
$\langle \hat{\rho}\hat{\rho} \rangle$. 
To close the equation  we approximate
$\langle \hat{\rho}\hat{\rho} \rangle \sim \langle \hat{\rho}\rangle
 \langle \hat{\rho} \rangle$. This leads to:
\begin{equation}
{ {d \rho^{\sigma\sigma'}_{ij}} \over {dt} }
        = i \sum_k (t_{ki}\rho^{\sigma\sigma'}_{kj} 
        - \rho^{\sigma\sigma'}_{ik} t_{jk}) 
+ 2iU \sum_{\gamma}(\rho^{\sigma\gamma}_{ij} (\vec{m}_j \cdot 
        \vec{\tau}_{\sigma'\gamma}) 
         - (\vec{\tau}_{\gamma\sigma} \cdot \vec{m}_i) 
        \rho^{\gamma\sigma'}_{ij}) 
         + i \frac{U}{2} (n_i - n_j) 
        \rho^{\sigma\sigma'}_{ij}
\end{equation}
Where,
\begin{equation}
        n_i(t) = \sum_{\sigma} \rho_{ii}^{\sigma\sigma}(t),~~~
\vec{m}_{i}(t) = \langle \hat{\vec{s}}_i \rangle =
        \sum_{\sigma\sigma'} \vec{\tau}_{\sigma\sigma'} 
        \rho_{ii}^{\sigma\sigma'}(t)
\end{equation}
%
%

{\it Modeling the pump:}
The electrons in the system couple with the vector 
potential $\vec{A}(\vec{r},t)$ 
associated with the electromagnetic field of the laser pulse. 
This can be 
included in the electronic part of the Hamiltonian via 
Peierl's substitution which transforms the tight-binding 
hopping parameter  as
\begin{equation}
    t_{ij} = \Tilde{t}_{ij} e^{i\int_{\vec{R}_i}^{\vec{R}_j}
    \vec{A}(t)\cdot \vec{dr}}, ~~~
    \vec{E} = -\frac{\partial \vec{A}}{\partial t},~~~
\vec{E}(t) = \vec{E}_0 e^{-\frac{(t-t_0)^2}{2 \tau_p^2}}
    \text{sin}(\omega_p t)
\end{equation}
Where $\omega_p$ and $\tau_p$ are the frequency and 
the width of the pulse. Equipped with these, we solve the 
MFD equation using the 
RK4 algorithm. 
The direction of the field is kept at $45^{\circ}$ to the x-axis.

\begin{figure}[t!]
\centering{
\includegraphics[width=8cm,height=7cm]{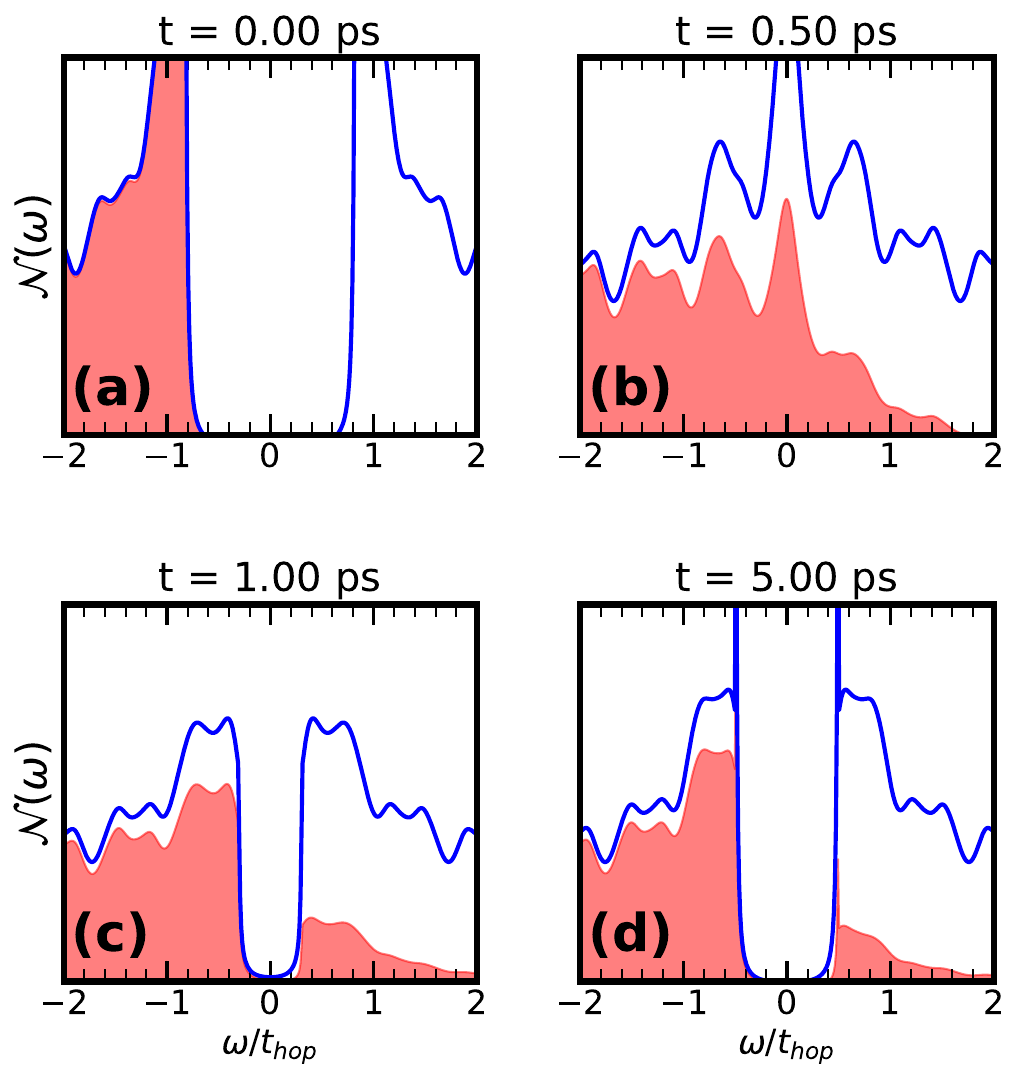}
\includegraphics[width=6cm,height=6.5cm]{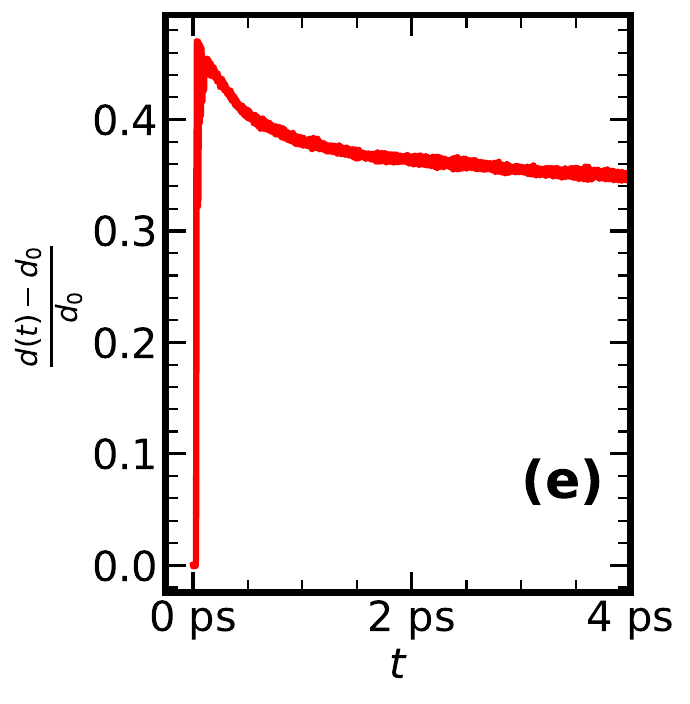}
}
\caption{(a)-(d): Instantaneous density of states $\mathcal{N}(\omega)$
[blue solid line] and the occupied part of the density of states
$g(\omega)\mathcal{N}(\omega)$ [color filled with red] shown for 
$E_0^2\sim 0.5$ at various times. (e) shows the double occupancy 
$d(t)$ as a function of time.
}
\end{figure}
\begin{figure}[b]
\centering{
\includegraphics[width=9cm,height=8cm]{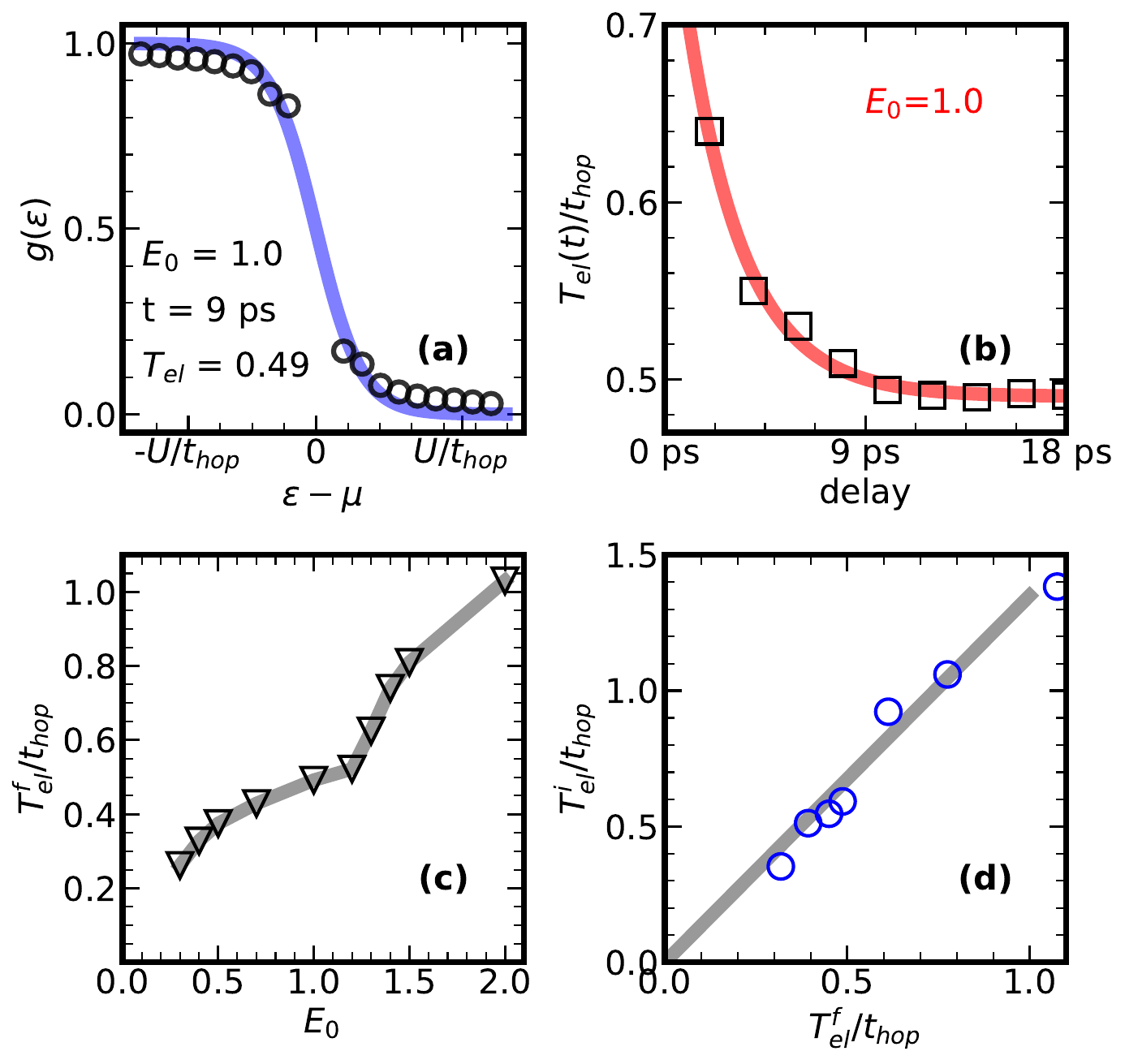}
}
\caption{Electronic temperature $T_{el}$ from 
MFD, by fitting the population function $g(\epsilon,t)$
on a $16\times16$ lattice with a 
Fermi function with temperature $T_{el}(t)$. 
(a)~A specific fit for pulse with
amplitude $E_0 =1.0$, at time t=9 ps.
(b)~$T_{el}(t)$ as obtained by fitting $g(\epsilon,t)$
at various times.
We behaviour can be fitted to
$T_{el}(t)=T_{el}^f + (T_{el}^i-T_{el}^f) e^{-t/\tau_{el}}$.
Independent of $E_0$,  $\tau_{el} \sim 3 $ps. 
(c)~$T_{el}^f$ as a function of $E_0$.
(d)~The relation between $T_{el}^i$ and
$T_{el}^f$ for changing $E_0$. The ratio is $\sim 1.5$. 
This allows
us to parametrise the pump only in terms of
$T_{el}^f$.
}
\end{figure}
\begin{figure}[t]
\centering{
\includegraphics[width=10cm,height=13cm]{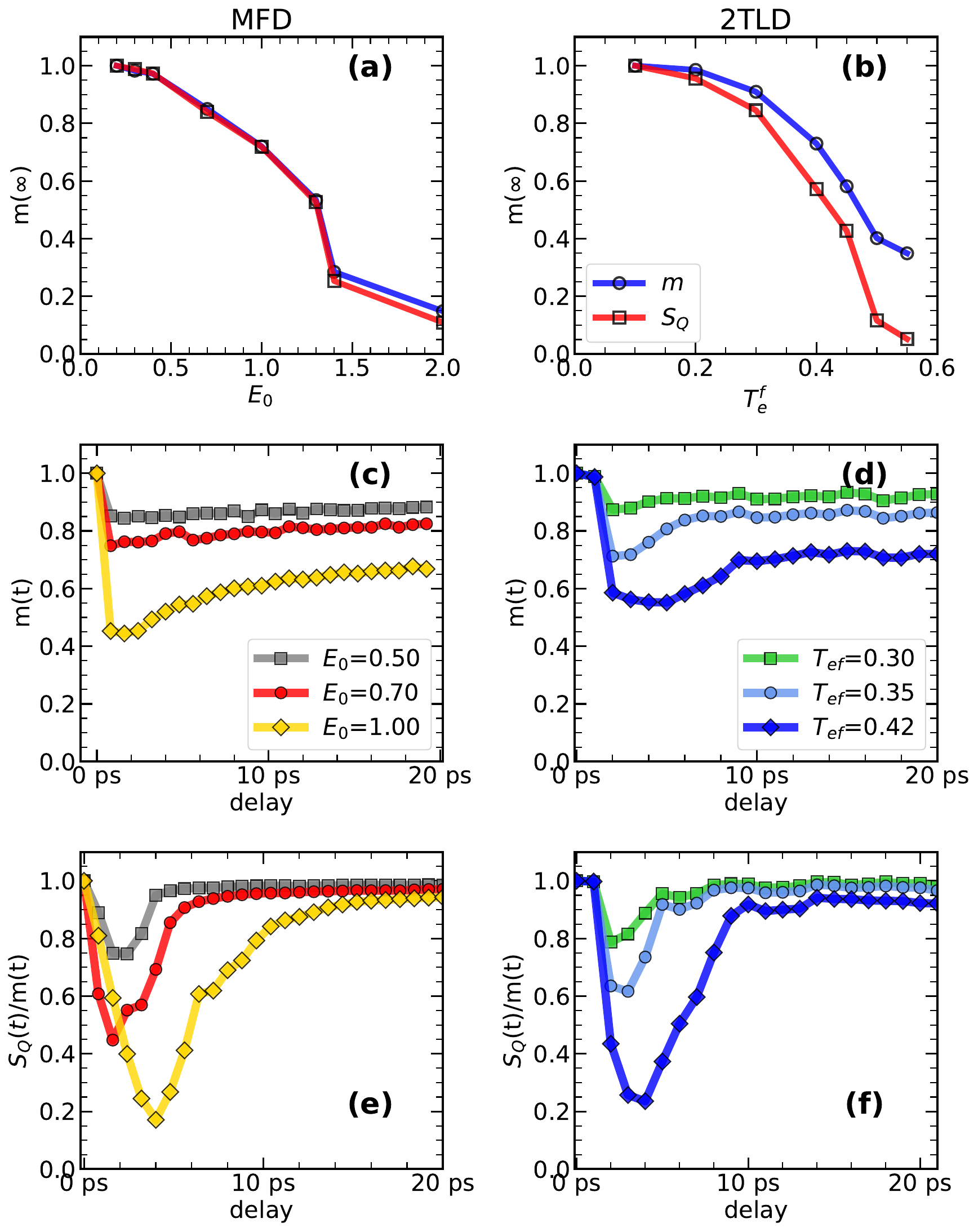}
}
\caption{
MFD results (left) can be captured by using a $T_{el}(t)$ in 
LD (right), with just one parameter $T_{el}^f$ replacing the fluence 
of the laser pulse.
The top panels (a, b) show the long-time magnetic moment and 
order parameter.
Middle panels (c, d) show $m(t)$, MFD and LD show 
similar decay and recovery with similar time scales.
Bottom panels (e, f) show $S_Q(t)/m(t)$ which separates the
order recovery from the moment recovery. This too compares
well. All these are for a single value of $\gamma$ in LD.
}
\end{figure}
\section{Supplement B: Electronic temperature from MFD}

We focus on the mean-field dynamics on a 
\(16 \times 16\) lattice with a pulse width of \(3/t_{\text{hop}}\). 
We maintain the pump frequency approximately equal to the
 gap in the density of states and vary only the amplitude 
\(E_0\). We calculate the instantaneous occupation 
 function $g(\epsilon,t)$:
\begin{equation}
g(\epsilon,t) \mathcal{N}(\epsilon,t) = \sum_{l} \langle 
f^{\dagger}_{l}(t) f_{l}(t) \rangle \delta (\epsilon - \epsilon_{l}(t)),~~~
\text{where}~~~
\mathcal{N}(\epsilon,t) = \sum_l \delta (\epsilon - \epsilon_{l}(t))
\end{equation}

Here, \(\epsilon_l\) are the eigenvalues in the instantaneous
background at time $t$. The
 operator \(f_l\) is the \(l\)-th instantaneous annihilation 
 operator at time \(t\), which annihilates an electron in the
  state corresponding to \(\epsilon_l\) in the background field.
The density of states and the occupied part of the density of states
are shown in Fig 1(a)-(d). In Fig.1(e) we plot the time dependence 
of the double occupancy, calculated from the instantaneous
magnetic background. The double occupancy $d(t)$ at half filling
can be estimated from the average square of the magnetic 
moment size as $d = \frac{1}{4}  - \langle m_i^2 \rangle$, 
where the average is over all the sites.

While $g(\epsilon,t)$ initially has a `non thermal' look it 
quickly takes a form that can be approximated by a Fermi 
distribution with
electronic temperature $T_e(t)$. 
Fig.2 shows such a fit.
We parameterize the extracted 
$T_e(t)$ with an exponential decay:
\begin{equation}
    T_{el}(t)=T_{el}^f+(T_{el}^i-T_{el}^f) e^{-t/\tau_{el}}
\end{equation}
where $T_{el}^i$ is the initial electron temperature 
(as soon as the pulse passes)
and $T_{el}^f$ is the long time electron temperature. 
We find that
$\tau_{el}$ is mainly fluence independent. 
From these we can establish a relationship between 
$E_0$ and $T_{el}^i, ~T_{el}^f$.  We find $T_{el}^i/T_{el}^f \sim 
1.5$ as shown in Fig.2., though the initial distribution cannot be
fitted with an Fermi-distribution reliably.

\section{Supplement C: Benchmarking Langevin with MFD}

One can write a Langevin-like stochastic equation [1], where 
the origin of the damping is phenomenological. We adapt this
formulation to 
calculate the effective torque arising from the 
excited electron population.  The spins experience a lower 
temperature, $T_b$. We call this scheme
2-Temperature Langevin Dynamics 
(2TLD) [Discussed in the main text].
On a lattice size of $16 \times 16$, we run 2TLD with the $T_{el}(t)$
obtained from the MFD. 
We used various values of the dissipation coefficient $\gamma$ 
in the LD to match LD results with MFD. If we want to single
$\gamma$ for all $E_0$ then $\gamma=0.1$ seems to be a
reasonable choice, however it requires a larger value of $T_{el}^i 
\sim 2.5 T_{el}^f$.  We compare the transient dynamics from 
MFD and LD in Fig.3.

In Fig.3 top panels 
we plot the steady-state value of $m$ and $S_{\textbf{Q}}$ as 
a function of $E_0$ obtained from both methods. In the middle 
panel, we plot $m(t)$, which shows similar quick suppression 
followed by a sub-10-ps recovery. In the bottom panels, we 
plot $S_{\textbf{Q}}(t)/m(t)$. 

As this is a very small system, the domain dynamics is severely 
size-dependent at this lattice length. Nonetheless, the ratio
$S_{\textbf{Q}}(t)/m(t)$ captures only the domain recovery, 
ignoring the effect of recovery of $m(t)$ on $S_{\textbf{Q}}(t)$. 
We see that the 2TLD method is reasonable within the range we
want to capture for larger system sizes compared to the MFD. 
We keep $T_b$ small, $\sim 0.01 t_{hop}^2/U$, for 2TLD.

\section{Supplement D: Averaging and $L$ dependence in LD}

The order recovery process is highly stochastic due to 
the role of domain growth. Therefore, it is necessary 
to average the recovery over multiple thermal runs. 
We used 5 to 10 different runs to average $S_{\textbf{Q}}(t)$. 
In Fig.4, we present the different traces as well as 
the averaged curve. Although the recovery timescale for 
the average magnetic moment $m(t)$ does not significantly 
depend on the system size, the recovery timescale 
for $S_{\textbf{Q}}(t)$ is size-dependent, as shown in
Fig.5.

 \begin{figure}[t]
 \centering{
 \includegraphics[width=8cm,height=5cm]{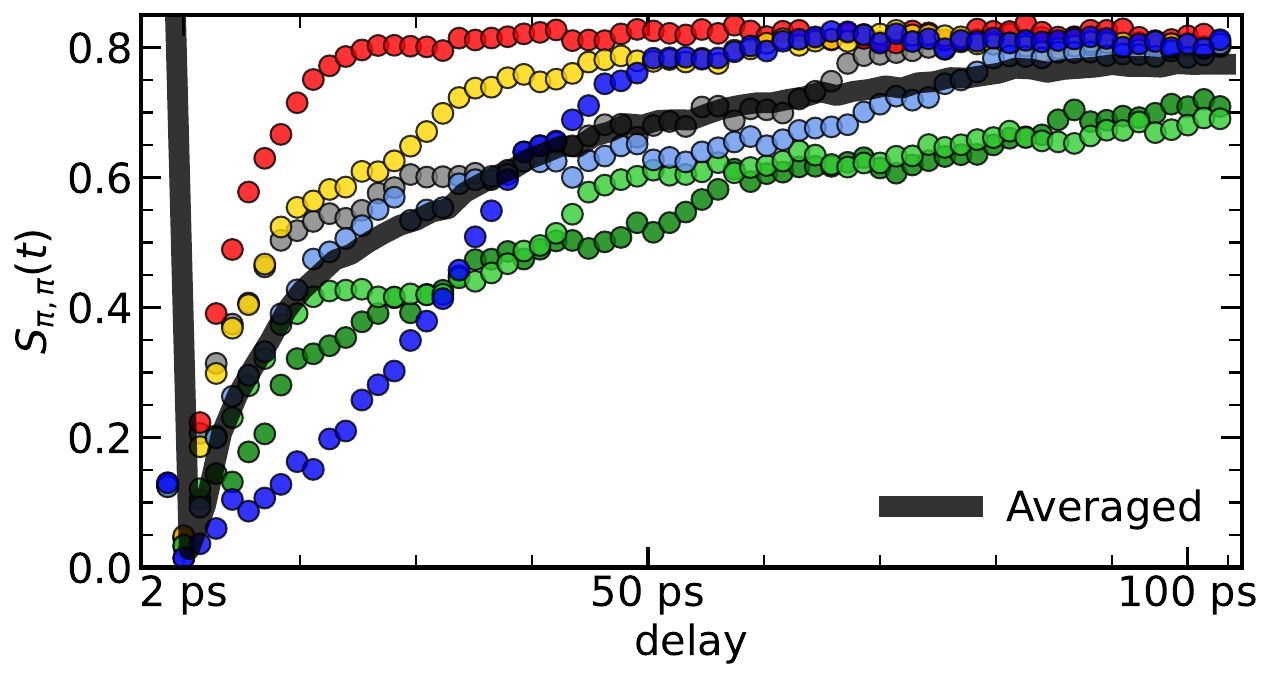}
 }
 \caption{Stochastic recovery of long range order:
 we plot the $S_{Q}(t)$ for different runs at
 $E_0^2=0.2$.
 The black solid line shows the average over
different runs. }
 \end{figure}

\begin{figure}[t]
\centering{
\includegraphics[width=12cm,height=6cm]{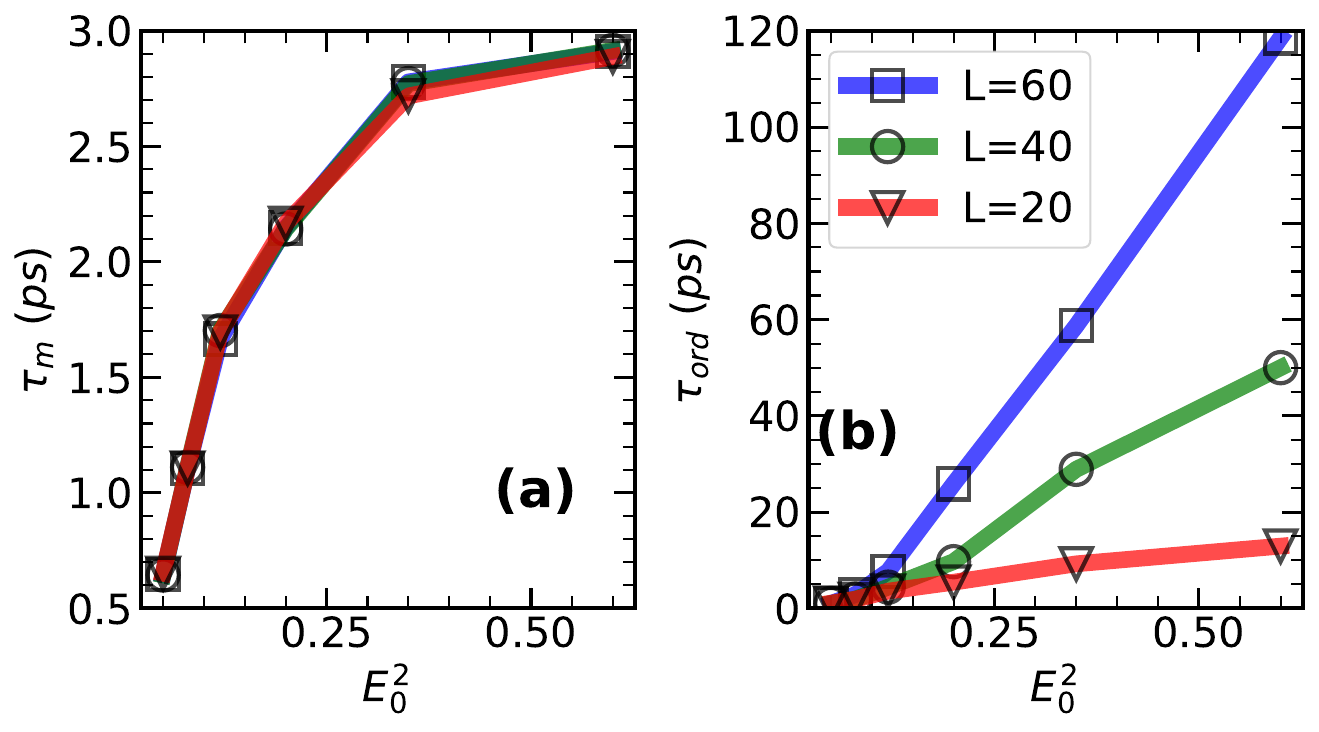}
}
\caption{Size dependence: The size dependence of the growth
time scale $\tau_m$ (a) and $\tau_{ord}$ (b) are plotted for system 
size L= 20, 40 and 60.
$\tau_m$ shows no dependence on the system size but 
$\tau_{ord}$ grows with system size.
}
\end{figure}
\begin{figure}[b!]
\includegraphics[width=16cm,height=5cm]{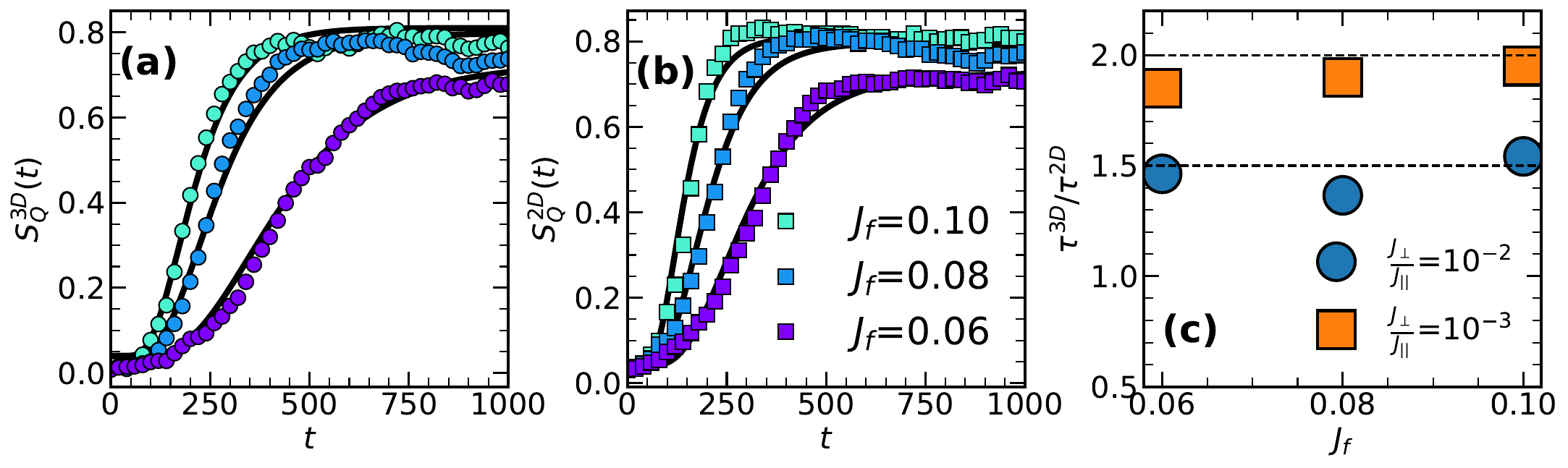}
\caption{Exploring a quasi-2D Heisenberg model with $J$
 for in-plane coupling and weak inter-plane coupling 
 $J_{\perp} \sim 10^{-3} J$. Using a $30 \times 30 \times 10$ lattice,
  we compute 3D and 2D structure factors, averaging over layers. 
  The structure factor growth was fitted with $S_{\textbf{Q}}(t)\sim
   e^{-(\tau/t)^{\alpha}}$, $\alpha\sim 1-2$. Despite $J_{\perp}$ being 
   1000 times weaker, the ratio of the recovery timescales
    ($\tau^{3D}/\tau^{2D}$) is only $\sim 2$.}
\end{figure}
\section{Supplement E: Optical conductivity calculation }
To calculate the optical conductivity, we assume a separation 
of time scales between charge and spin dynamics. For an 
instantaneous background ${\vec{m}_i}$, we diagonalize the single 
particle Hamiltonian as described in the main text. This yields 
the eigenfunctions ($f_{i,\epsilon}$) and eigenvalues ($\epsilon_n$). 
The current operator is defined as $j_{mn}^x = 
\langle m|\hat{j}_x|n\rangle$. The electron temperature is set 
to $T_{el}(t)$. The conductivity is then computed
as follows:
\begin{equation}
\sigma(\omega) = \sum_{m \neq n}
\frac{|j_{mn}|^2}{\epsilon_m - \epsilon_n}
\delta(\omega - (\epsilon_n - \epsilon_m)) [f_{\epsilon_m}(T_{el}) -
f_{\epsilon_n}(T_{el})],
\end{equation}
where $f_{\epsilon}(T)$ denotes the Fermi function. A more 
rigorous expression for conductivity, using two-time correlation,
is provided in [2].
\vspace{1cm}

\section{Supplement F: Recovery in the Heisenberg Model}

At half-filling and large $U/t$ Hubbard model 
maps to the Heisenberg model.
\begin{equation}
    H_{heis} = \sum_{ij} J_{ij} \vec{S}_i \cdot \vec{S}_j,
\end{equation}
where $J_{ij} = \frac{t^2_{ij}}{U} m^2$, with $m$ being the average 
magnetic moment (magnitude of 1 at equilibrium) and $\vec{S}_i$ a 
unit vector in $SO(3)$. In a layered system the in-plane $J_{||}$
exchange is $\gg$ the interlayer exchange  $J_{\perp}$.
The spin dynamics follows the Landau–Lifshitz–Gilbert-Brown 
(LLGB) equation:
\begin{equation}
    \frac{d\vec{S}_i}{dt} = - \vec{S}_i \times 
    \left(\frac{\partial H_{heis}}{\partial \vec{S}_i} + 
    \vec{\eta}_i(t)\right) + \gamma \vec{S}_i \times \frac{d\vec{S}_i}{dt},
\end{equation}
where $\vec{\eta}_i$ represents thermal noise from a 
bath at temperature $T_b$ [3]. 
Numerical method like the Suzuki-Trotter 
decomposition is employed to solve this equation 
[4].

{\it Initial condition:} After the pump pulse passes, the 
system is assumed to 
have a reduced magnetic moment and a distorted 
spin configuration. 
The evolution parameters are $\gamma$, 
$T_b$, $J_{||}$, and $J_{\perp}$.

To analyze the effect of photo-pumping on recovery dynamics, 
we make the following approximations:
\begin{itemize}
    \item The initial state corresponds to $S_{\textbf{Q}} 
    \sim 0$ and is derived from a thermal state with $T_b > T_N$ 
    (where $T_N$ is the critical temperature).
    \item The system evolves with nearest-neighbor in-plane 
    coupling $J_{f}$ and dissipation rate $\gamma$.
\end{itemize}
{\it Parameters:} We set $J_{||} = J_f$ and keep the
 $J_{\perp}/J_{||}$ ratio constant 
and small ($10^{-3}$, $10^{-2}$). The bath temperature $T_b$ is 
maintained at $0.4 \times T_N$, and $\gamma$ is set to 0.05. 
We use a 3D lattice of dimensions $30 \times 30 \times 10$ 
(10 layers in the $z$-direction).
We compute the 2D structure factor $S_{\textbf{Q}}^{2D}(t)$ 
averaged 
over all layers and the 3D structure factor 
$S_{\textbf{Q}}^{3D}(t)$. 
Starting with an initial state where all layers have
 $S_{\textbf{Q}}^{2D}\sim 0$,
the system evolves with varying $J_f$.
Five parallel runs were averaged for the 3D 
structure factor growth. 

{\it 2D vs 3D recovery:}
Fig.6 shows the detailed dynamics. Both 2D and 
3D structure factor 
growths are fitted with $S_{\bf Q} = S_{\bf Q}^0
 e^{-(\tau_{rec}/t)^{\alpha}}$, 
and we extract two timescales, $\tau^{2D}$ and $\tau^{3D}$, 
plotting their ratio with $J_f$. Although the
 inter-layer coupling 
is 1000 times weaker than the in-plane coupling, 
the recovery timescale is only a factor of
 approximately 2 times larger.

{\it Role of dissipation:}
Fig.7 shows the dependence of $S_{\bf Q}^{2D}$
 on $\gamma$.
With $\gamma$ values ranging from 0.01 to 0.05. 
Our results indicate that $\tau_{rec} \propto J_f
 \times \frac{1}{\gamma ^{3/4}}$.

\begin{figure}[t]
\centering{
\includegraphics[width=12cm,height=6cm]{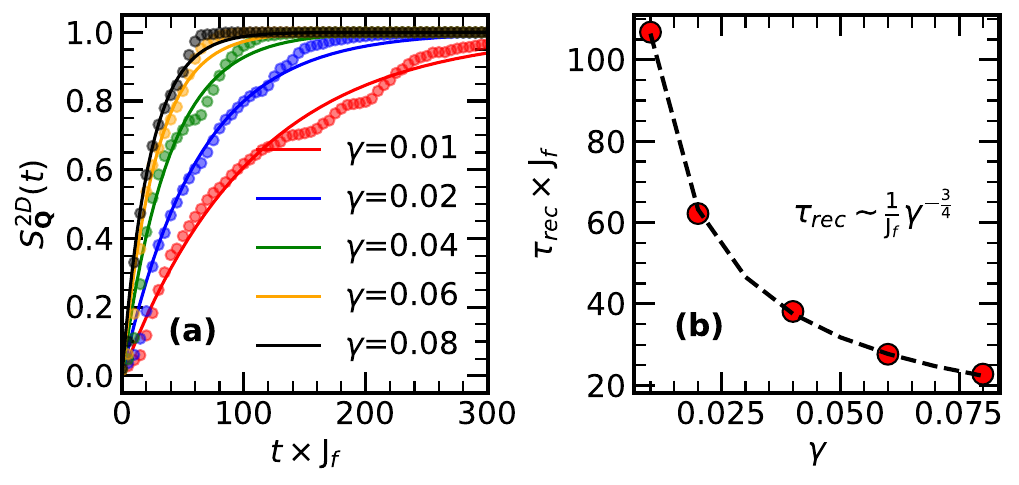}
}
\caption{(a) $S_{\textbf{Q}}^{2D}(t)$ for the
 same initial condition 
(thermal state with $T > T_N$) but different
 dissipation rates $\gamma$. 
Solid lines represent fits. (b) Dependence 
of $\tau_{rec}$ on $\gamma$, 
showing a relationship: $\tau_{rec} \propto 
J_f \times \frac{1}{\gamma ^{3/4}}$.}
\end{figure}

\vspace{2em}

\noindent \textbf{The bibliography}

\vspace{1em}

[1] Pui-Wai Ma and S. L. Dudarev,  
Longitudinal magnetic fluctuations in Langevin spin dynamics,  
Phys. Rev. B 86, 054416 (2012),  
\href{https://doi.org/10.1103/PhysRevB.86.054416}{DOI: 10.1103/PhysRevB.86.054416}

\vspace{1em}

[2] Zala Lenarčič, Denis Golež, Janez Bonča, and Peter Prelovšek,  
Optical response of highly excited particles in a strongly correlated system,  
Phys. Rev. B 89, 125123 (2014),  
\href{https://doi.org/10.1103/PhysRevB.89.125123}{DOI: 10.1103/PhysRevB.89.125123}
\vspace{1em}

[3] Sauri Bhattacharyya, Sankha Subhra Bakshi, Saurabh Pradhan, and Pinaki Majumdar,  
Strongly anharmonic collective modes in a coupled electron-phonon-spin problem,  
Phys. Rev. B 101, 125130 (2020),  
\href{https://doi.org/10.1103/PhysRevB.101.125130}{DOI: 10.1103/PhysRevB.101.125130}

\vspace{1em}

[4] Pui-Wai Ma and S. L. Dudarev,  
Langevin spin dynamics,  
Phys. Rev. B 83, 134418 (2011),  
\href{https://doi.org/10.1103/PhysRevB.83.134418}{DOI: 10.1103/PhysRevB.83.134418}


\begin{thebibliography}{100} 
\bibitem{Grant}
P. M. Grant, S. S. P. Parkin, V. Y. Lee, E. M. Engler, M. L. Ramirez, J. E. Vazquez, G. Lim, R. D. Jacowitz, and R. L. Greene,
{\it Evidence for superconductivity in La$_2$CuO$_4$},
\href{https://doi.org/10.1103/PhysRevLett.58.2482}{Phys. Rev. Lett. 58, 2482 (1987)}.

\bibitem{lee-nag-rmp}
Patrick A. Lee, Naoto Nagaosa, and Xiao-Gang Wen, 
{\it Doping a Mott insulator: Physics of high-temperature superconductivity},
\href{https://doi.org/10.1103/RevModPhys.78.17}{Rev. Mod. Phys. 78, 17 (2006)}.

\bibitem{Imada}
Masatoshi Imada, Atsushi Fujimori, and Yoshinori Tokura,
{\it Metal-insulator transitions}
\href{https://doi.org/10.1103/RevModPhys.70.1039}{Rev. Mod. Phys. 70, 1039 – 1998}.

\bibitem{Perdew}
J. P. Perdew and Alex Zunger,
{\it Self-interaction correction to density-functional approximations 
for many-electron systems}, 
\href{https://doi.org/10.1103/PhysRevB.23.5048}{Phys. Rev. B 23, 5048 – 1981}.

\bibitem{Anisimov}
Vladimir I. Anisimov, Jan Zaanen, and Ole K. Andersen,
{\it Band theory and Mott insulators: Hubbard U instead of Stoner I}
\href{https://doi.org/10.1103/PhysRevB.44.943}{Phys. Rev. B 44, 943 – 1991}.

\bibitem{review1}
Alberto de la Torre, Dante M. Kennes, Martin Claassen, Simon Gerber, James W. McIver, and Michael A. Sentef,
{\it Colloquium: Nonthermal pathways to ultrafast control in quantum materials},
\href{https://doi.org/10.1103/RevModPhys.93.041002}{Rev. Mod. Phys. 93, 041002 – 2021}

\bibitem{review2}
Yuta Murakami, Denis Golež, Martin Eckstein, Philipp Werner,
{\it Photo-induced nonequilibrium states in Mott insulators},
\href{https://doi.org/10.48550/arXiv.2310.05201}{arXiv:2310.05201 [cond-mat.str-el]}

\bibitem{review3}
Hideo Aoki, Naoto Tsuji, Martin Eckstein, Marcus Kollar, Takashi Oka, and Philipp Werner,
{\it Nonequilibrium dynamical mean-field theory and its applications},
\href{https://doi.org/10.1103/RevModPhys.86.779}{Rev. Mod. Phys. 86, 779 – 2014}

\bibitem{ppmott1}
Martin Eckstein and Philipp Werner, {\it Photoinduced States in a Mott Insulator}
\href{https://doi.org/10.1103/PhysRevLett.110.126401}{Phys. Rev. Lett. 110, 126401 – 2013}

\bibitem{ppmott2}
Takashi Oka and Hideo Aoki, {\it Photoinduced Tomonaga-Luttinger-like liquid in a Mott insulator}
\href{https://doi.org/10.1103/PhysRevB.78.241104}{Phys. Rev. B 78, 241104(R) – 2008}

\bibitem{ppmott3}
Zhuoran He and Andrew J. Millis
{\it Photoinduced phase transitions in narrow-gap Mott insulators: The case of VO$_2$}
\href{https://doi.org/10.1103/PhysRevB.93.115126}{Phys. Rev. B 93, 115126 – 2016}

\bibitem{ppmott4}
Takashi Oka and Hideo Aoki,
{\it Photoinduced Tomonaga-Luttinger-like liquid in a Mott insulator}
\href{https://doi.org/10.1103/PhysRevB.78.241104}{Phys. Rev. B 78, 241104(R) – 2008}

\bibitem{ppmott5}
Jiajun Li, Markus Müller, Aaram J. Kim, Andreas M. Läuchli, and Philipp Werner,
{\it Twisted chiral superconductivity in photodoped frustrated Mott insulators}
\href{https://doi.org/10.1103/PhysRevB.107.205115}{Phys. Rev. B 107, 205115 – 2023}

\bibitem{ppmott6}
M. Ligges, I. Avigo, D. Golež, H.U.R. Strand, Y. Beyazit, K. Hanff, F. Diekmann, L. Stojchevska, M. Kalläne, P. Zhou, K. Rossnagel, M. Eckstein, P. Werner, and U. Bovensiepen,
{\it Ultrafast Doublon Dynamics in Photoexcited $1T$-TaS$_2$},
\href{https://doi.org/10.1103/PhysRevLett.120.166401}{Phys. Rev. Lett. 120, 166401 – 2018}


\bibitem{ppmott7}
Takashi Oka,
{\it Nonlinear doublon production in a Mott insulator: Landau-Dykhne method applied to an integrable model},
\href{https://doi.org/10.1103/PhysRevB.86.075148}{Phys. Rev. B 86, 075148 – 2012}

\bibitem{ppmott8}
Ryota Ueda, Kazuhiko Kuroki, and Tatsuya Kaneko,
{\it Photoinduced $\eta$-pairing correlation in the Hubbard ladder},
\href{https://doi.org/10.1103/PhysRevB.109.075122}{Phys. Rev. B 109, 075122 – 2024}

\bibitem{ppmott9}
Julián Rincón, Elbio Dagotto, and Adrian E. Feiguin,
{\it Photoinduced Hund excitons in the breakdown of a two-orbital Mott insulator},
\href{https://doi.org/10.1103/PhysRevB.97.235104}{Phys. Rev. B 97, 235104 – 2018}

\bibitem{ppmott10}
Jiajun Li and Martin Eckstein,
{\it Nonequilibrium steady-state theory of photodoped Mott insulators},
\href{https://doi.org/10.1103/PhysRevB.103.045133}{Phys. Rev. B 103, 045133 – 2021}

\bibitem{ppmott11}
K. Kimura, H. Matsuzaki, S. Takaishi, M. Yamashita, and H. Okamoto,
{\it Ultrafast photoinduced transitions in charge density wave, Mott insulator, and metallic phases of an iodine-bridged platinum compound},
\href{https://doi.org/10.1103/PhysRevB.79.075116}{Phys. Rev. B 79, 075116 – Published 18 February 2009}

\bibitem{ppmott12}
Eckstein, M., Werner, P. {\it Ultra-fast photo-carrier relaxation in Mott insulators with short-range spin correlations}, \href{https://doi.org/10.1038/srep21235}{Sci Rep 6, 21235 (2016)}

\bibitem{ppmott13}
Akira Takahashi, Hisashi Itoh, and Masaki Aihara,
{\it Photoinduced insulator-metal transition in one-dimensional Mott insulators},
\href{https://doi.org/10.1103/PhysRevB.77.205105}{Phys. Rev. B 77, 205105 – 2008}

\bibitem{Zala}
Zala Lenarčič and Peter Prelovšek, {\it Ultrafast Charge 
Recombination in a Photoexcited Mott-Hubbard Insulator}, 
\href{https://doi.org/10.1103/PhysRevLett.111.016401}{Phys. Rev. Lett. 111, 016401 – 2013}.

\bibitem{Huang}
T.-S. Huang, C. L. Baldwin, M. Hafezi, and V. Galitski, 
{\it Spin-mediated Mott excitons}, 
\href{https://doi.org/10.1103/PhysRevB.107.075111}{Phys. Rev. B 107, 
075111 – 2023}.

\bibitem{Wrobel}
P. Wróbel and R. Eder,{\it Excitons in Mott insulators}, 
\href{https://doi.org/10.1103/PhysRevB.66.035111}{Phys. Rev. B 66, 035111 – 2002}.

\bibitem{material1}
B. J. Kim et.al. 
{\it Novel $J_{eff}=1/2$ Mott State Induced by Relativistic Spin-Orbit 
Coupling in $Sr_2 IrO_4$}, 
\href{https://doi.org/10.1103/PhysRevLett.101.076402}{PRL 101, 076402 (2008)}.

\bibitem{material2}
C. L. Lu, J.-M. Liu, {\it The J$_{eff}$ = 1/2 Antiferromagnet 
Sr$_2$IrO$_4$: A Golden Avenue toward New Physics and 
Functions}. Adv. Mater. 2020, 32, 1904508
\href{https://doi.org/10.1002/adma.201904508}{Adv. Mater. 2020, 32, 1904508}.

\bibitem{material3}
R. Arita, J. Kuneš, A. V. Kozhevnikov, A. G. Eguiluz, and M. Imada,
{\it Ab initio Studies on the Interplay between Spin-Orbit Interaction 
and Coulomb Correlation in Sr$_2$IrO$_4$ and Ba$_2$IrO$_4$}, 
\href{https://doi.org/10.1103/PhysRevLett.108.086403}{Phys. Rev. Lett. 108, 086403 – 2012}.


\bibitem{material4}
Li, Q., Cao, G., Okamoto, S. et al. {\it Atomically resolved 
spectroscopic study of Sr$_2$IrO$_4$: Experiment and theory}. 
\href{https://doi.org/10.1038/srep03073}{Sci Rep 3, 3073 (2013)}.

\bibitem{experiment1}
Dean, M., Cao, Y., Liu, X. et al.
{\it Ultrafast energy- and momentum-resolved
dynamics of magnetic correlations in the
photo-doped Mott insulator $Sr_2 IrO_4$},
\href{https://doi.org/10.1038/nmat4641}{Nature Mater 15, 601–605 (2016)}.

\bibitem{experiment2}
Mehio, O., Li, X., Ning, H. et al. 
{\it A Hubbard exciton fluid in a photo-doped antiferromagnetic 
Mott insulator.} 
\href{https://doi.org/10.1038/s41567-023-02204-2}{Nat. Phys. 19, 1876–1882 (2023)}.

\bibitem{ED1}
Akira Takahashi, Hisashi Itoh, and Masaki Aihara, 
{\it Photoinduced insulator-metal transition in one-dimensional 
Mott insulators}, 
\href{https://doi.org/10.1103/PhysRevB.77.205105}{Phys. Rev. B 77, 205105 – 2008}.

\bibitem{ED2}
Satoshi Ejima, Florian Lange, and Holger Fehske,
{\it Photoinduced metallization of excitonic insulators},
\href{https://doi.org/10.1103/PhysRevB.105.245126}{Phys. Rev. B 105, 245126 – 2022}

\bibitem{DMFT1}
Afanasiev, Gatilova, et.al.
{\it Ultrafast Spin Dynamics in Photodoped Spin-Orbit Mott 
Insulator $Sr_2IrO_4$}
\href{https://doi.org/10.1103/PhysRevX.9.021020}{Phys. Rev. X 9, 021020 (2019)}.

\bibitem{DMFT2}
P. Werner, N. Tsuji, and M. Eckstein, 
{\it Nonthermal Symmetry-Broken States in the Strongly Interacting 
Hubbard Model},
\href{https://doi.org/10.1103/PhysRevB.86.205101}{Phys. Rev. B 86, 205101 (2012)}.

\bibitem{DMFT3}
J. H. Mentink and M. Eckstein, {\it Ultrafast Quenching of the
Exchange Interaction in a Mott Insulator}, 
\href{https://doi.org/10.1103/PhysRevLett.113.057201}{Phys. Rev. Lett. 113, 057201 (2014)}.

\bibitem{DMFT4}
K. Balzer, F. A. Wolf, I. P. McCulloch, P. Werner, and M.
Eckstein, {\it Nonthermal Melting of Néel Order in the Hubbard
Model}, 
\href{https://doi.org/10.1103/PhysRevX.5.031039}{Phys. Rev. X 5, 031039 (2015)}.

\bibitem{DMFT5}
M. Eckstein and P. Werner, {\it Ultra-fast Photo-Carrier
Relaxation in Mott Insulators with Short-Range Spin
Correlations}, 
\href{https://doi.org/10.1038/srep21235}{Sci. Rep. 6, 21235 (2016)}.

\bibitem{DMFT6}
J. K. Freericks, V. M. Turkowski, and V. Zlatić, 
{\it Nonequilibrium Dynamical Mean-Field Theory}
\href{https://doi.org/10.1103/PhysRevLett.97.266408}{Phys. Rev. Lett. 97, 266408 – 2006}

\bibitem{DMRG1}
Satoshi Ejima, Florian Lange, and Holger Fehske,
{\it Nonequilibrium dynamics in pumped Mott insulators},
\href{https://doi.org/10.1103/PhysRevResearch.4.L012012}{Phys. Rev. Research 4, L012012 – 2022}.

\bibitem{DMRG2}
Satoshi Ejima, Florian Lange, and Holger Fehske,
{\it Photoinduced metallization of excitonic insulators},
\href{https://doi.org/10.1103/PhysRevB.105.245126}{Phys. Rev. B 105, 245126 – 2022}.

\bibitem{MFD}
Chern, Gia-Wei and Barros, Kipton et.al.
{\it Semiclassical dynamics of spin density waves},
\href{https://doi.org/10.1103/PhysRevB.97.035120}{PhysRevB.97.035120, (2018)}.

\bibitem{suppl}
All sections are included in the supplementary material. 
In Supplement A, we derive the mean-field 
equation of motion for equal time correlations under 
an external electric 
pulse. In Supplement B, we establish the relationship 
between the electric 
field and effective electronic temperature. Supplement C 
compares the 
dynamics from MFD and non-equilibrium Langevin dynamics. 
In Supplement D, 
we discuss the effect of system size. Supplement E provides
the formula we used to calculated the conductivity.
Supplement F compares 
the 3D recovery dynamics to the layer-averaged 2D recovery 
of the structure 
factor in a quasi-2D Heisenberg model with 
$J_{\perp}/J_{||}\sim 10^{-3}$.

\bibitem{heis1}
Bang-Gui Liu, 
{\it A nonlinear spin-wave theory of quasi-2D quantum
Heisenberg antiferromagnets}
J. Phys.: Condens. Matter 4 8339 (1992)

\bibitem{heis2}
A. Du and G. Z. Wei, 
{\it Magnetic Properties of Layered Heisenberg Ferromagnets},
Aust. J. Phys., 1993, 46, 571-81.

\bibitem{therm}
Rajdeep Sensarma, David Pekker, Ehud Altman, Eugene Demler,
Niels Strohmaier, Daniel Greif, Robert Jördens, Leticia Tarruell, 
Henning Moritz, and Tilman Esslinger
{\it Lifetime of double occupancies in the Fermi-Hubbard model}
\href{https://doi.org/10.1103/PhysRevB.82.224302}{Phys. Rev. B 82, 224302 – (2010)}

\end{thebibliography}
\end{document}